\documentclass[sigconf]{acmart}

\AtBeginDocument{%
  }

\setcopyright{acmlicensed}
\copyrightyear{2025}
\acmYear{2025}
\setcopyright{cc}
\setcctype{by}
\acmConference[UIST '25]{The 38th Annual ACM Symposium on User Interface
Software and Technology}{September 28-October 1, 2025}{Busan, Republic of
Korea}
\acmBooktitle{The 38th Annual ACM Symposium on User Interface Software and
Technology (UIST '25), September 28-October 1, 2025, Busan, Republic of
Korea}
\acmDOI{10.1145/3746059.3747602}
\acmISBN{979-8-4007-2037-6/2025/09}

\acmSubmissionID{8878}

\begin{document}

\title{\textit{Understood}: Real-Time Communication Support for Adults with ADHD Using Mixed Reality}

\author{Shizhen Zhang}
\authornote{Both authors contributed equally to this research.}
\email{zhangshzh@shanghaitech.edu.cn}
\orcid{0009-0003-6704-0170}
\affiliation{%
  \institution{School of Information Science and Technology, ShanghaiTech University}
  \city{Shanghai}
  \country{China}
}

\author{Shengxin Li}
\authornotemark[1]
\email{shengxinli2002@gmail.com}
\orcid{0009-0006-4117-3954}
\affiliation{%
  \institution{School of Information Science and Technology, ShanghaiTech University}
  \city{Shanghai}
  \country{China}
}

\author{Quan Li}
\authornote{Corresponding Author.}
\email{liquan@shanghaitech.edu.cn}
\orcid{0000-0003-2249-0728}
\affiliation{%
  \institution{School of Information Science and Technology, ShanghaiTech University}
  \city{Shanghai}
  \country{China}
}

\renewcommand{\shortauthors}{Shizhen and Shengxin et al.}

\begin{abstract}
  Adults with Attention Deficit Hyperactivity Disorder (ADHD) often experience communication challenges, primarily due to executive dysfunction and emotional dysregulation, even after years of social integration. While existing interventions predominantly target children through structured or intrusive methods, adults lack tools that translate clinical strategies into daily communication support. To address this gap, we present \textit{Understood}, a Mixed Reality (MR) system implemented on \textit{Microsoft HoloLens 2}, designed to assist adults with ADHD in real-world communication. Through formative semi-structured interviews and a design workshop, we identified critical communication barriers and derived design goals for the system. \textit{Understood} combines three key features: (1) \textit{real-time conversation summarization} to reduce cognitive load, (2) \textit{context-aware subsequent word suggestions} during moments of disfluency, and (3) \textit{topic shifting detection and reminding} to mitigate off-topic transitions. A within-subjects user study and expert interviews demonstrate that \textit{Understood} effectively supports communication with high usability, offering a complement to therapist-mediated interventions.

\end{abstract}

\begin{CCSXML}
<ccs2012>
   <concept>
       <concept_id>10003120.10011738.10011776</concept_id>
       <concept_desc>Human-centered computing~Accessibility systems and tools</concept_desc>
       <concept_significance>500</concept_significance>
       </concept>
 </ccs2012>
\end{CCSXML}

\ccsdesc[500]{Human-centered computing~Accessibility systems and tools}

\keywords{Attention Deficit Hyperactivity Disorder (ADHD), Mixed Reality, Communication Support, Computer Mediated Communication}


\begin{teaserfigure}
 \includegraphics[width=\textwidth]{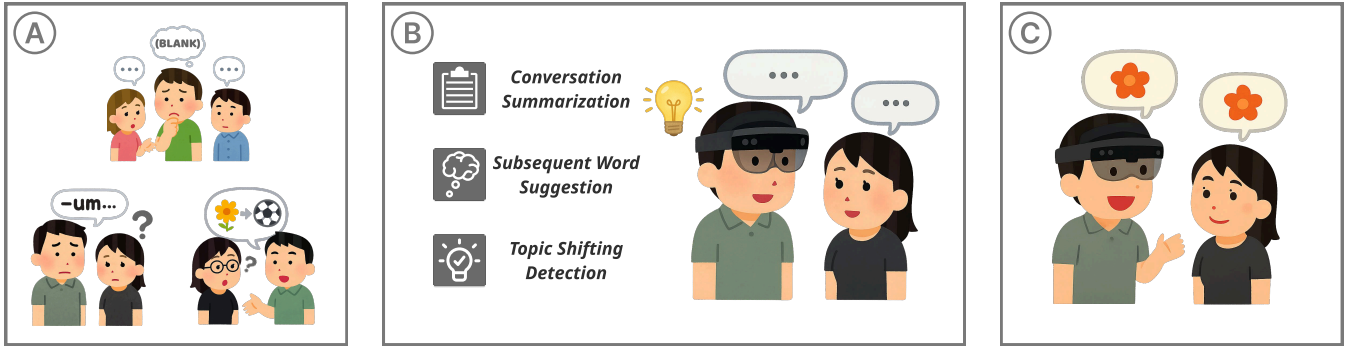}
 \caption{Example scenario using \textit{Understood}. (A) Individuals with ADHD may face difficulties navigating real-world communication scenarios, often struggling with attention regulation and conversational flow. (B) To seek support, the user adopts the \textit{HoloLens} headset and activates \textit{Understood}. The system provides real-time conversation summarization, context-aware subsequent word suggestions, and topic shifting detections and remindings to help manage the discourse. (C) With the assistance of \textit{Understood}, the user engages in the conversation with greater confidence and fluency, interacting more effectively with communication partners.}
 \Description{Three-panel illustration labeled A, B, and C showing a scenario involving communication challenges and assistive technology. (1) Panel A: Depicts several individuals experiencing communication difficulties. One person has a thought bubble labeled “(BLANK),” another says “um…” while looking confused, and a third person shows jumbled thoughts involving unrelated icons like a flower and a soccer ball. (2) Panel B: Shows a person wearing a HoloLens headset conversing with another person. Icons on the left indicate system features: conversation summarization, subsequent word suggestion, and topic shifting detection. (3) Panel C: The same two individuals converse smoothly, both smiling with matching flower icons in their speech bubbles, indicating improved mutual understanding.}
 \label{fig:teaser}
\end{teaserfigure}

\received{20 February 2007}
\received[revised]{12 March 2009}
\received[accepted]{5 June 2009}

\maketitle

\section{Introduction}

\par Attention-Deficit Hyperactivity Disorder (ADHD) is a persistent neurodevelopmental disorder marked by patterns of inattention, hyperactivity, and impulsivity that significantly impair daily functioning~\cite{kooij2010european}. Epidemiological studies estimate that ADHD affects $3\%$-$10\%$ of school-aged children and $1\%$-$6\%$ of adults worldwide, with approximately one to two thirds of childhood cases continuing into adulthood, necessitating lifelong adaptation strategies~\cite{wender2001adults}. While diagnostic criteria have traditionally focused on the core behavioral symptoms in children~\cite{katsarou2024identifying,mathers2006aspects,martinussen2006working,mikami2015social,storebo2019social,wilkes2017pragmatic,wenchen2016language}, recent research increasingly highlights the characteristics and impact of ADHD in adults~\cite{surman2013understanding,safren2006cognitive,solanto2011cognitive,wender2001adults,barkley2007adhd-ch13}. Specifically, ADHD in adults presents distinct phenotypes: whereas hyperactivity is more salient in childhood and often facilitates earlier diagnosis, adulthood manifestations tend to shift toward emotional dysregulation and executive dysfunction, reflecting more complex and internalized symptoms.~\cite{katzman2017adult}. Notably, adults with ADHD experience marked impairments in executive functions~\cite{boonstra2005executive,spencer2008triple}, working memory~\cite{ossmann2003inhibition}, and cognitive flexibility~\cite{roshani2020comparison}, which together contribute to the broader difficulties associated with the disorder. These neural and behavioral differences collectively contribute to dysfunction across multiple life domains.

\par The impact of these differences becomes particularly evident in interpersonal communication~\cite{westby2021adhd}. Clinical studies demonstrate that many individuals with ADHD report persistent social communication challenges~\cite{wenchen2016language,theodoratou2024communication,mathers2006aspects}, despite these difficulties not being included in the core diagnostic criteria~\cite{american2013diagnostic}. These communication difficulties are rooted in the neurobiological and behavioral characteristics of ADHD and manifest in distinct patterns across various interaction contexts. Clinical evaluations indicate that individuals with ADHD often experience ``Stuckness'' during conversations, where they struggle to maintain a fluid dialogue~\cite{katsarou2024identifying,tannock2018adhd}, exhibit impaired discourse cohesion with tangential narratives~\cite{baker1992attention}, and disrupt conversational turn-taking with premature responses~\cite{bellani2011language,mathers2006aspects}. Collectively, these challenges create a ``conversational paradox'' — individuals with ADHD often possess strong verbal intelligence, yet their executive dysfunction impedes effective message delivery.

\par Current non-medicinal interventions have shown limited efficacy in addressing the real-time communication challenges inherent in ADHD management. While evidence-based behavioral interventions~\cite{collins2005strategy,reimers1987acceptability}, including Cognitive Behavioral Therapy (CBT)~\cite{solanto2011cognitive,safren2006cognitive}, Social Skills Training (SST)~\cite{mikami2017social,storebo2019social,mikami2015social}, and Self-Regulation Training~\cite{barkley2004attention,surman2013understanding} exhibit clinical effectiveness in controlled settings, their applicability in real-world communication remains constrained. For instance, SST promotes structured behavioral modifications, but acquired skills often fail to generalize beyond simulated scenarios~\cite{willis2019stand}. CBT targets maladaptive thought patterns, yet its reliance on cognitive flexibility limits its real-time adaptability~\cite{antshel2014cognitive}. Similarly, self-regulation training enhances awareness of emotional states but does not sufficiently support spontaneous, socially appropriate responses~\cite{cibrian2022potential}. Technology-integrated supports present potential solutions: VR systems provide immersive, controlled exposure~\cite{wong2023effectiveness}, but often foster passive, prompted learning; Wearable biofeedback devices aim to improve regulation through brain activity training~\cite{swingle2008biofeedback,doren2019sustained}, though concerns remain regarding their intrusiveness and the lack of regulatory oversight, which may leads to unreliable outcomes~\cite{geppert2019neuro}. Additionally, prior work in HCI, mainly with children, has demonstrated autonomy-supportive system designs for ADHD~\cite{stefanidi2023children,silva2023unpacking,sonne2016changing}, but these remain limited in addressing real-time communication needs in adults.

\par The above landscape reveals three key gaps: \textbf{G1: Sustainable transfer of therapeutic outcomes} - current systems rely on retrospective analysis, restricting skill-building to clinical contexts and making it difficult to maintain behavioral improvements in natural social interactions; \textbf{G2: Empowerment of user autonomy in ADHD adults} - Although much of the prior work, particularly with children, has supported my in system-mediated interventions for ADHD, existing implementations often still treat ADHD adults as passive diagnostic subjects, rather than empowering them to actively regulate their own communicative patterns; and \textbf{G3: System with minimal user disruption} - many existing solutions disrupt the user's natural conversational flow through overt notifications and cognitively demanding interactions. Our work seeks to bridge these gaps by integrating computational linguistics, wearable devices, and neurodiversity-informed design.

\par Therefore, our study seeks to explore the following four research questions (\textbf{RQ1--RQ4}). To begin, we aim to understand communication challenges faced by individuals with ADHD and identify suitable support strategies. This leads us to two foundational research questions: \textbf{RQ1:} \textit{What communication challenges do individuals with ADHD encounter, and what are the current intervention strategies?} and  \textbf{RQ2:} \textit{What strategies could better support individuals with ADHD in mitigating these communication challenges?} To investigate \textbf{RQ1}, we conducted two semi-structured interview studies involving four licensed clinicians and 12 adults with ADHD. These interviews revealed three primary barriers: (1) \textit{deficits in working memory}; (2) \textit{disfluency and pauses in conversation}; and (3) \textit{attentional anchoring failure (e.g., abrupt topic shifts)}. In addressing \textbf{RQ2}, we carried out a design study involving clinicians, researchers, and ADHD people, that began with an exploratory design brainstorming session to generate diverse concepts targeting these communication issues. We then developed storyboards to visualize the concepts and held a design workshop to further refine the supportive approaches. MR emerged as a preferred medium due to its potential to deliver information non-intrusively, minimize conversational disruptions, and support polite interaction.

\par Based on feedback from clinicians, researchers, and ADHD people, we defined five design goals and developed \textit{Understood}, an MR system implemented on \textit{Microsoft HoloLens 2}\footnote{\url{https://learn.microsoft.com/en-us/hololens/}}, designed to support adults with ADHD in real-world communication. \textit{Understood} incorporates three core features: \textbf{(1) real-time conversation summarization} to address working memory limitations and reduce cognitive load, \textbf{(2) context-aware subsequent word suggestions} explicitly activated by users during moments of disfluency or hesitation, and \textbf{(3) topic shift detection and reminding} to help mitigate off-topic transitions.

\par In this study, we further explored two key research questions concerning the \textit{Understood} system: \textbf{RQ3:} \textit{How effectively does Understood support communication?} and \textbf{RQ4:} \textit{What is the usability of Understood?} To answer these questions, we conducted a within-subjects study in which participants completed communication tasks under two conditions: (1) using \textit{Understood} and (2) without it. We collected both quantitative performance metrics and qualitative feedback through semi-structured interviews, enabling a comprehensive evaluation of the system's communication support (RQ3) and usability (RQ4). The findings consistently demonstrated that when evaluated under a within-subject, lab-based setting, \textit{Understood} significantly improved communication fluency for individuals with ADHD while maintaining a high level of usability. In summary, our contributions are as follows:
\begin{itemize} 
    \item Conducted semi-structured interviews with clinicians and individuals with ADHD to identify communication challenges and define the system's support needs;
    \item Performed a design study involving clinicians, researchers, and individuals with ADHD to explore the design space and iteratively refine the design of an accessible, therapist-free, real-time support system; 
    \item Developed the MR system, \textit{Understood}, designed to assist adults with ADHD in real-world communication contexts; 
    \item Conducted a within-subjects user study to systematically assess the effectiveness and usability of \textit{Understood}, while also identifying design trade-offs.
\end{itemize}

\section{Background and Related Work}
\par In this section, we begin by providing a theoretical overview of ADHD and the communication challenges faced by individuals with ADHD. We then review existing support strategies aimed at addressing these challenges, focusing on (1) conventional behavioral interventions and (2) emerging technology-integrated supports. This background and analysis of current approaches help identify key gaps and inform the development of more effective, tailored systems for individuals with ADHD.

\subsection{ADHD and Communication Challenges}

\par ADHD is a neurodevelopmental condition characterized by persistent patterns of inattention, impulsivity, and hyperactivity, which disrupt various aspects of daily life~\cite{wender2001adults,furman2005attention}. These symptoms are particularly challenging in conversational settings, where individuals with ADHD often face difficulties with both verbal and non-verbal communication. These challenges arise from the core symptoms of the condition, manifesting in distinctive ways during interactions and hindering their ability to engage effectively in interpersonal communication~\cite{wenchen2016language,theodoratou2024communication}.

\par One significant challenge observed in individuals with ADHD is a phenomenon known as ``Stuckness'' during conversations~\cite{katsarou2024identifying}. This occurs when a person becomes mentally fixed on a specific topic, word, or thought, making it difficult to transition smoothly or adapt to the natural flow of dialogue~\cite{tannock2018adhd}. As a result, conversations may feature long pauses, repeated phrases, or abrupt topic shifts, often leaving conversational partners confused or disengaged. Another communication barrier is incoherent logic in discourse~\cite{baker1992attention}. Individuals with ADHD often struggle to organize their thoughts, leading to fragmented or tangential speech patterns. They might jump between unrelated ideas or fail to provide sufficient context, making it difficult for others to grasp their intended message. This disorganization is closely linked to impairments in working memory and executive functioning -- cognitive processes responsible for planning, organizing, and regulating thoughts ~\cite{martinussen2006working,mcinnes2003listening}. Additionally, inappropriate wording and poor self-regulation in speech are common, with impulsivity often leading to interruptions, blurting out responses, or using contextually inappropriate language~\cite{bellani2011language,mathers2006aspects}. These behaviors can cause misunderstandings, disrupt conversational flow, and create social friction. Together, these challenges present significant obstacles to effective communication in both personal and professional scenarios.

\par Given that many of these difficulties stem from deficits in executive functions, such as working memory, cognitive flexibility, and inhibitory control~\cite{bellani2011language}, our study aims to enhance conversational fluency, cohesion, and overall social interactions for individuals with ADHD by addressing these specific communication challenges. To achieve this, we propose a system that offers real-time feedback and guidance, supporting individuals with ADHD during their conversations.

\begin{figure*}[h]
    \centering
    \includegraphics[width=\linewidth]{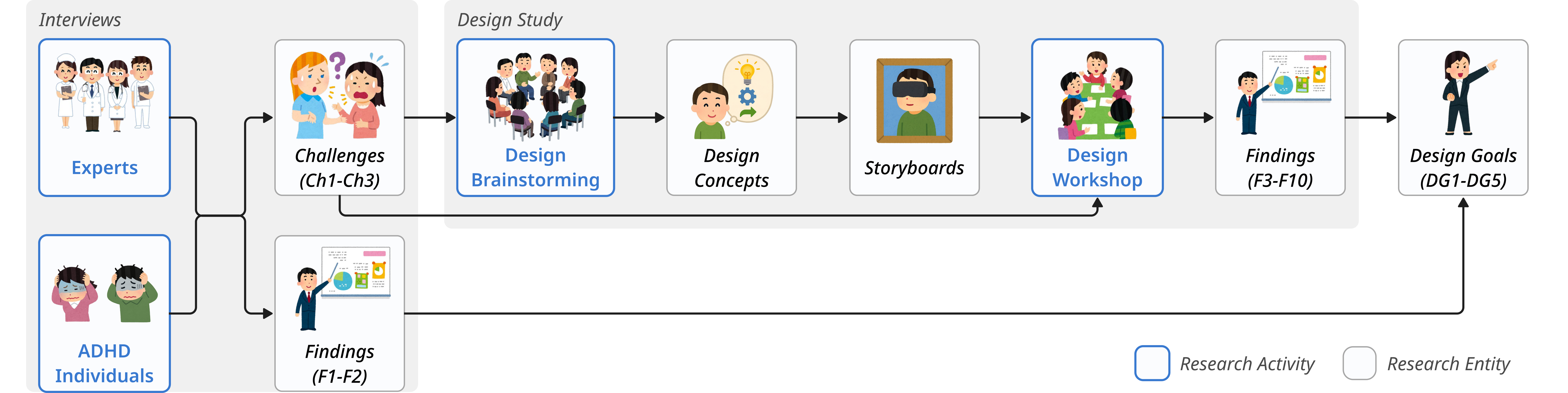}
    \caption{An overview of the formative study procedure. The blue boxes are different research activities while the gray boxes indicate research entities.}
    \label{fig:FormativeStudyProcedure}
    \Description{A flowchart diagram showing the formative study procedure, with blue boxes representing research activities and gray boxes representing research entities. The process starts with two groups under "Interviews": "Experts" (illustrated as medical professionals) and "ADHD Individuals" (illustrated as people appearing distressed). Arrows lead to identified "Challenges (Ch1–Ch3)" and initial "Findings (F1–F2)". These feed into a "Design Study" phase, including: "Design Brainstorming" (group discussion), "Design Concepts" (a person with a lightbulb and gears), "Storyboards" (a person wearing a HoloLens), and a "Design Workshop" (group activity). The workshop produces more "Findings (F3–F10)," leading to final "Design Goals (DG1–DG5)," depicted as a person confidently pointing forward.}
\end{figure*}

\subsection{Behavioral Interventions for Communication Challenges in ADHD}

\par Effectively addressing communication challenges in individuals with ADHD requires interventions that enhance fluency and coherence during conversations. The conventional non-medical approach primarily relies on \textbf{behavioral interventions}~\cite{collins2005strategy,reimers1987acceptability}, which focus on developing foundational skills through structured training~\cite{rajeh2017interventions,daley2014behavioral,caye2019treatment}. These interventions typically target areas such as active listening~\cite{clayton2021behavioral}, staying on topic~\cite{anderson1991improving}, and self-regulating speech~\cite{graham1992developing}.

\par Behavioral interventions have traditionally been the primary approach to supporting individuals with ADHD in improving communication skills. These interventions focus on developing strategies to manage impulsivity, enhance attention, and organize thoughts during conversations~\cite{theodoratou2024communication,wilkes2017pragmatic}. One key method is \underline{CBT}~\cite{solanto2011cognitive,safren2006cognitive}, which aims to alter maladaptive thought patterns that disrupt communication. By utilizing self-monitoring and cognitive restructuring, individuals learn to identify and manage behaviors such as distractibility and impulsivity. Antshel et al.~\cite{antshel2014cognitive} found that CBT can effectively improve conversational skills and self-regulation in individuals with ADHD, although its success is contingent on the individual's engagement and consistent application of strategies, it also requires cognitive flexibility and real-time self-monitoring, which are often impaired in adults with ADHD. Another widely-used method is \underline{SST}~\cite{mikami2017social,storebo2019social,mikami2015social}, which targets specific conversational competencies such as turn-taking, initiating and maintaining dialogue, and interpreting social cues. Through role-playing exercises, individuals practice real-world scenarios to improve their social interactions. Research by Willis et al.~\cite{willis2019stand} suggests that SST can reduce social isolation and improve communication, though its benefits may remain confined to structured sessions and fail to generalize to spontaneous, real-world conversations, where working memory and cognitive flexibility are crucial. Additionally, \underline{Self-regulation training}~\cite{barkley2004attention,surman2013understanding} incorporates techniques like mindfulness to help individuals control impulsive speech and emotional responses during conversations. Studies, such as those by Cibrian et al.~\cite{cibrian2022potential}, show that self-regulation training can improve attention and emotional responses in ADHD, particularly in challenging social situations. However, while it enhances emotional awareness, this does not necessarily translate into timely or socially appropriate responses during real-time interactions. Moreover, this approach requires significant time and effort to achieve lasting results and may be less effective if ADHD symptoms remain unmanaged.

\par Despite their benefits, these interventions are typically structured and may not easily transfer to spontaneous, real-time conversations. Their success also heavily depends on the individual's commitment, making long-term adherence a challenge. These limitations highlight the need for complementary solutions that can provide real-time support in dynamic conversational contexts.

\subsection{Technology-Integrated Supports for Communication Challenges}

\par Researchers are increasingly developing technology-integrated supports that use computational methods to analyze communication difficulties in ADHD and provide interactive support~\cite{wong2023effectiveness,raeisian2025confident,Sehlin2018experiences,keshav2019digital}. These emerging technologies serve as valuable extensions to conventional interventions by offering real-time feedback and dynamic assistance during conversations. For example, Virtual Reality (VR) technologies offer immersive environments for practicing communication in realistic settings. Wong et al.~\cite{wong2023effectiveness} developed a VR-based system designed to help individuals with ADHD practice social skills. While these systems provide a safe, controlled environment for practice, their practical implementation remains limited due to the constrained range of scenarios they provide. Some of these solutions combine wearable devices with biofeedback, or specifically neurofeedback systems to monitor physiological indicators, such as heart rate, skin conductance, and brain activity~\cite{swingle2008biofeedback,doren2019sustained,geppert2019neuro}. Swingle et al.~\cite{swingle2008biofeedback} demonstrated that, through training, individuals can learn to modify their brainwave patterns to better manage cognitive challenges such as inattention and impulsivity, thereby indirectly improving communication skills. However, despite demonstrated effectiveness, the application of these methods in clinical settings often deviates from research-based protocols due to insufficient regulation, leading to inconsistent outcomes and reduced reliability. Some research has also explored how system design can support autonomy in individuals with ADHD~\cite{stefanidi2023children,silva2023unpacking,sonne2016changing}. Stefanidi et al.~\cite{stefanidi2023children} proposed approaches that move beyond symptom-oriented interventions by engaging the broader care ecosystem. However, these efforts have primarily targeted children, with limited attention to autonomy-supportive communication tools for adults with ADHD.

\par In addition to research focused on ADHD, several studies have explored communication support for individuals with various communication challenges~\cite{zulfikar2024memoro,bermejo2020vimes,hohenstein2023artificial,ou2024academic,elsahar2019augmentative,yu2024holoaac,curtis2024looking}. Elsahar et al.~\cite{elsahar2019augmentative} provided a comprehensive review of tools designed to assist individuals with speech and language impairments, evaluating them in terms of accessibility, affordability, complexity, portability, and typical conversation speed. Similarly, Hohenstein et al.~\cite{hohenstein2023artificial} and Ou et al.~\cite{ou2024academic} investigated how artificial intelligence shapes users’ perceptions and influences linguistic and social interactions in both everyday and academic settings. Regarding specific tools, Curtis et al.~\cite{curtis2024looking} evaluated three high-fidelity prototypes—embody projection, worn audio devices, and headset-driven MR -- through the lens of visual modalities, highlighting the potential of current and emerging MR technologies in enhancing communication. Zulfikar et al.~\cite{zulfikar2024memoro}, on the other hand, focused on auditory modalities and propose a wearable memory-augmentation system. Although these works do not specifically target individuals with ADHD, they offer valuable design insights that can inform our own implementation.

\par Building upon these foundations, our work emphasizes low-intrusiveness interaction paradigms that seamlessly integrate real-time support while preserving the natural flow of conversation. By providing instant, context-specific feedback during face-to-face interactions, our system helps individuals with ADHD manage conversational flow, regulate emotions, and improve logical coherence without disrupting the conversation's continuity. This approach prioritizes timely intervention to address challenges as they occur, ultimately fostering more effective and confident communication.

\section{Formative Study}
\par Our formative study adopts a neurodiversity-affirming perspective to explore communication challenges among individuals with ADHD and identify supportive strategies through a two-stage process. First, semi-structured interviews were conducted to uncover key communication difficulties and user needs. Building on these insights, we engaged in a participatory design study involving individuals with ADHD, HCI researchers, and clinicians to co-develop potential solutions and establish design goals for a communication support system. The procedure of the formative study can be seen at \autoref{fig:FormativeStudyProcedure}.

\subsection{Semi-Structured Interviews}
\label{sec:3.1}

\par To address \textbf{RQ1}, we conducted semi-structured interviews with two key stakeholder groups: four licensed clinicians and $12$ adults diagnosed with ADHD. By integrating clinical expertise with firsthand lived experiences, this dual-perspective methodology enabled a holistic analysis of communication barriers and intervention efficacy, yielding multifaceted insights that directly guided system requirements and design priorities.

\subsubsection{Expert Interview}
\label{sec:ExpertInterview}

\par We conducted semi-structured interviews with four licensed clinicians specializing in ADHD (\textbf{C1--C4}): two psychiatrists (\textbf{C1--C2}) and two therapists (\textbf{C3--C4}). Participants were recruited from local hospitals and had an average of 11.75 years of experience (SD = 2.86). To be eligible, clinicians were required to have managed at least 50 ADHD cases and to have directly observed clients' natural communication patterns. Each interview lasted approximately 60 minutes, followed IRB guidelines, and participants received a compensation of \$20 USD.

\par The interview protocol began with an exploration of communication challenges associated with ADHD, utilizing 15 clinically validated statements on pragmatic communication skills derived from the \textit{Diagnostic Interview for ADHD in Adults (DIVA-5)}~\cite{kooij2019diva}, a widely used diagnostic tool based on DSM-V criteria~\cite{american2013diagnostic}. Representative items included: \textit{``Has difficulty concentrating on a conversation,''} \textit{``Often changes the subject of the conversation,''} and \textit{``Needs many words to express a simple idea.''} All the statements can be seen at \autoref{app:DIVA-5}. Clinicians quantitatively assessed the prevalence of each challenge using a 3-point Likert scale (1 = \textit{uncommonly seen}, 3 = \textit{frequently seen}), followed by qualitative elaboration through anonymized case examples and analysis of contextual factors influencing these communication patterns. Additionally, we conducted an open-ended evaluation of existing interventions, where participants shared the strategies they employed to support individuals with ADHD in overcoming communication challenges, discussing both their advantages and limitations.

\subsubsection{ADHD Individuals Interview}
\par We conducted semi-structured interviews with 12 individuals diagnosed with ADHD (\textbf{P1-P12}; 5 male, 7 female; Mean age = 21.5 years, SD = 2.10), all of whom met the DSM-V criteria and were recruited from the local university. Each interview lasted approximately 25 minutes, followed IRB guidelines, and participants received a compensation of \$10 USD. The interviews began by exploring participants' communication challenges. To facilitate reflection on past experiences, we presented the same 15 DIVA-5 statements used in the Expert Interviews (\autoref{sec:ExpertInterview}). Participants indicated whether they had encountered the challenges described in each statement and were encouraged to share additional communication challenges not covered by the statements. For each endorsed challenge, they elaborated on specific situations where these difficulties arose. Additionally, participants were invited to discuss how they managed or coped with these challenges in their daily lives.

\subsubsection{Data Analysis}
\par To holistically map communication challenges and intervention opportunities, we conducted an integrated analysis of interview data from both clinicians and individuals with ADHD.

\par All sessions were video-recorded, transcribed, and analyzed using affinity diagramming techniques~\cite{harboe2015real,lucero2015using}, facilitated via \textit{Miro}\footnote{\url{https://miro.com/app/dashboard/}}. To ensure analytical rigor, the research team engaged in weekly consensus meetings to interpret findings and iteratively refine thematic categories. Cross-referencing clinician insights with ADHD people narratives on the collaborative Miro board enabled triangulation of emergent patterns, particularly behavioral manifestations of the 15 DIVA-5 communication statements observed in video recordings (e.g., topic shifts, disfluency). To validate findings, participants completed a member-checking process, reviewing preliminary results to annotate alignment or discrepancies with their experiences. This iterative feedback loop ensured fidelity to participant perspectives while refining the final analytical framework.

\subsubsection{Contextual Analysis Findings}
\par Based on the contextual analysis, we identified three key communication challenges (\textbf{Ch1-Ch3}) and two critical system requirements (\textbf{F1-F2}) to support individuals with ADHD.

\begin{figure}[h]
    \centering
    \includegraphics[width=\linewidth]{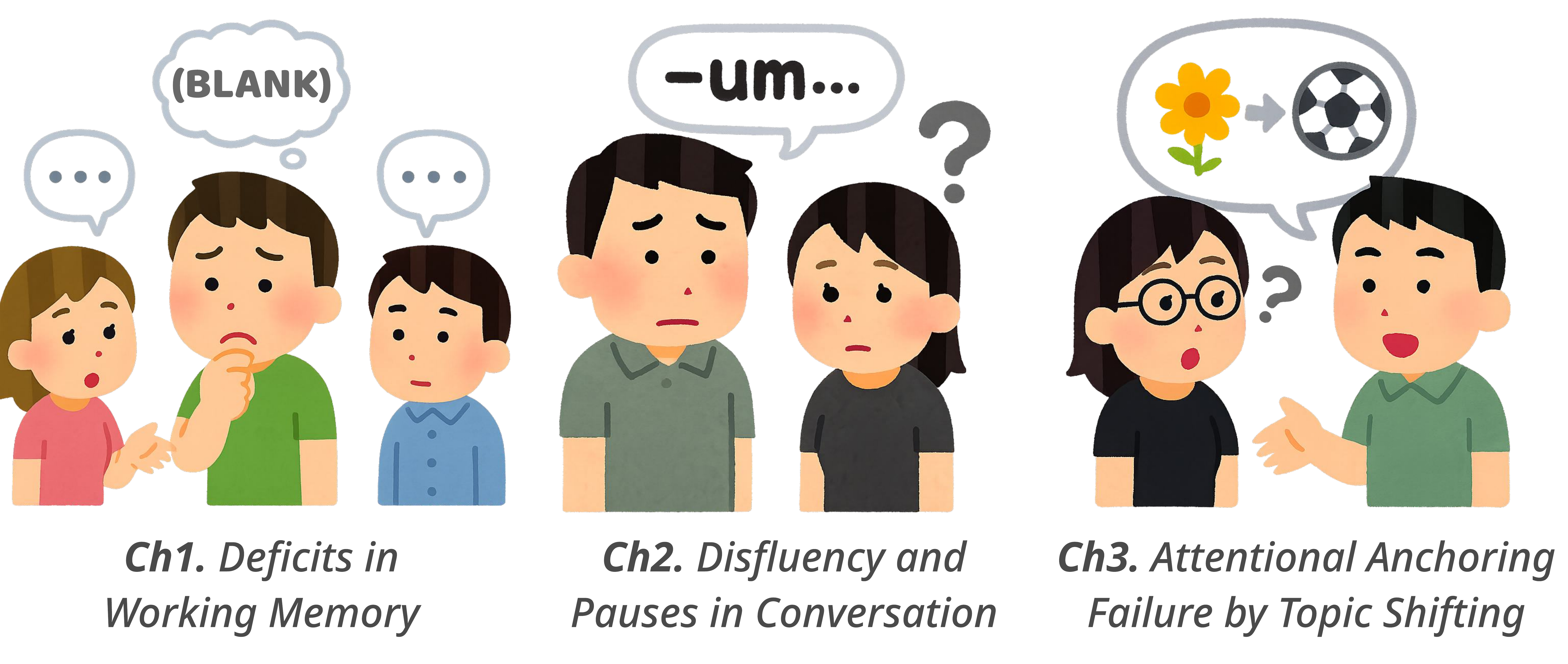}
    \caption{An overview of communication challenges encountered by individuals with ADHD, as found from the semi-structured interviews.}
    \label{fig:CommunicationChallenges}
    \Description{Illustration summarizing three communication challenges faced by individuals with ADHD. We use the same 3 illustrations as in the teaser panel A. (1) Ch1: Deficits in Working Memory – A person appears confused with a thought bubble labeled “(BLANK),” while two others look concerned. (2) Ch2: Disfluency and Pauses in Conversation – Two individuals look uneasy; one says “um…” with a question mark above their head, suggesting a struggle to maintain conversational flow. (3) Ch3: Attentional Anchoring Failure by Topic Shifting – A person talks while pointing, with icons of a flower and soccer ball above their head, while the listener looks puzzled, indicating difficulty staying focused when the topic changes.}
\end{figure}

\par \paragraph{Communication Challenges.} Insights, as illustrated in \autoref{fig:CommunicationChallenges}, were derived from both clinician observations and firsthand accounts of adults with ADHD, revealing recurring barriers in everyday interactions, and pinpointing actionable opportunities for technological interventions.

\par \textbf{Ch1: Deficits in Working Memory.} In response to \textit{\#6. I have difficulty filtering and/or selecting information}, all four clinicians rated this issue as \textbf{3} (frequently observed), and ten individuals with ADHD reported experiencing similar difficulties. Many described struggling to integrate new information with existing knowledge, often forgetting earlier content mid-conversation. This challenge was widely attributed to working memory deficits. As \textbf{P3} explained, \textit{``When others are speaking, it's hard for me to process their words. I can only remember a few of them, and I often miss important parts. I sometimes need to ask for clarification before I can respond properly.''}

\par All twelve individuals with ADHD acknowledged limitations in working memory, linking them to broader communication challenges. Clinicians echoed this view. As \textbf{C1} explained, \textit{``Working memory is the ability to hold and process relevant information. It's essential for understanding context and making inferences. Deficits here contribute to communication breakdowns.''} Some also noted that such deficits hindered sustained attention during conversations (\textit{\#3}).

\par \textbf{Ch2: Disfluency and Pauses in Conversation.} In response to \textit{\#2. I have difficulty speaking fluently in long sentences}, three clinicians rated this issue as \textbf{3} (frequently observed), and eight individuals with ADHD reported similar struggles. Two key underlying causes emerged from the interviews. The first was working memory deficits, where ADHD people described losing track of their thoughts mid-sentence. As \textbf{P5} put it, \textit{``When I try to express something, it's like... like... (stuck)... my mind just goes blank, and I have to pause to recover.''}

\par The second cause involved cognitive overload. Individuals with ADHD reported that too many thoughts surfaced simultaneously, making it difficult to organize speech. As \textbf{P9} shared, \textit{``I often think of both relevant and irrelevant things at once. I need time to sort them out and find the right words, so it probably looks like I'm pausing for too long.''} Several participants noted that external cues helped them resume speaking. For instance, \textbf{P4} explained, \textit{``When someone guesses the word I'm stuck on, it's like my memory kicks back in. I can keep going after that, like they've unblocked me.''}

\par \textbf{Ch3: Attentional Anchoring Failure Reflected by Topic Shifting.} In response to \textit{\#4. I often change the subject of the conversation}, three clinicians rated this issue as \textbf{3} (frequently observed), and nine individuals with ADHD reported similar experiences. As \textbf{P8} recounted, \textit{``My group was discussing a project, and suddenly I started talking about what I had for breakfast. It completely derailed the conversation.''} Likewise, \textbf{P10} noted, \textit{``I often bring up both relevant and irrelevant topics,''} linking this to \textit{\#9} (talk too much) and \textit{\#11} (need a lot of words to say).
    
\par Interestingly, several participants observed that topic shifting was less frequent in serious or formal settings, where social expectations helped anchor attention. As \textbf{C3} explained, \textit{``In early sessions, topic shifting is rare due to unfamiliarity. But as we build rapport, conversations become more casual, and shifting topics becomes more frequent.''}

\par \paragraph{Needs for Supportive System.} Interviews with both clinicians and individuals with ADHD highlighted a strong demand for comprehensive system-level support, as existing intervention methods were often regarded as limited in scope and effectiveness. These insights informed the high-level requirements for our proposed system.

\par \textbf{F1: Limitations of Existing Intervention Methods.} As noted in the Introduction, current interventions for ADHD fall into two main categories: medical and non-medical. While medical treatments such as \textit{Concerta}~\cite{concerta} can have been shown to alleviate communication difficulties~\cite{maras2024effect}, their side effects remain a major concern. As \textbf{P2} shared, \textit{``When I took Concerta, my mind felt clearer, and communication improved. But it also made me tense, nauseous, and disrupted my sleep -- so I only take it when absolutely necessary.''} Given that adult ADHD is a lifelong condition, the long-term sustainability of medication use is often compromised by these adverse effects. While non-medical interventions were generally perceived as beneficial, participants highlighted their limited applicability in real-world scenarios. Strategies learned in structured environments -- such as CBT or SST -- often proved difficult to implement during actual conversations due to executive function deficits. As \textbf{P1} explained, \textit{``In real conversations, I just don't have the mental bandwidth to apply those strategies.''} SST, in particular, was critiqued for its lack of ecological validity. \textbf{C3} noted, \textit{``After socializing, individuals with ADHD may overanalyze their mistakes, which increases hesitation and shame in future interactions. Helping them transfer learned skills from training to real-life situations remains a key challenge.''}

\par \textbf{F2: The Need for Accessible, Therapist-Free, and Real-Time Support.} Interviews highlighted a pressing need for accessible, real-time support that operates independently of professional supervision. While ADHD presents persistent communication challenges, its comparatively milder clinical presentation -- relative to conditions like depression or mania -- often results in less attention and fewer dedicated resources. Only 6 of the 12 participants reported ever receiving structured interventions such as CBT or SST. As \textbf{P7} noted, \textit{``It's so hard for me to book a therapy appointment. I had to find some books and hope they would help.''} This lack of access underscores the value of scalable, automated solutions that provide consistent support without requiring therapist involvement. Participants also emphasized the potential of real-time, context-sensitive assistance. Unlike traditional interventions confined to clinical settings, in-situ support embedded in everyday interactions could bridge the gap between learning and application. As \textbf{C4} observed, \textit{``Providing assistance on-the-fly could turn real-world environments into training spaces, helping ADHD people build resilience and reduce stigma.''} Similarly, \textbf{P3} remarked, \textit{``Real-time assistance may help me apply what I've been learned immediately.''}

\par \paragraph{Minor Findings} In addition to the major themes, several minor findings emerged from the interviews. First, regarding the issue of \textit{\#12. I have difficulty waiting my turn during conversations}, all four clinicians rated this behavior as \textbf{3} (frequently observed), and 11 individuals with ADHD reported experiencing similar challenges. However, 10 of them noted that years of social experience had helped them develop self-regulation strategies to manage impulsive behaviors, such as interrupting or speaking out of turn (\textit{\#12--15}). As \textbf{C4} noted, \textit{``This is one of the main differences between adults and children with ADHD. We provide less training on interrupting behaviors for adults.''} Additionally, four participants mentioned difficulty maintaining eye contact during conversations, often due to stigma-related self-consciousness. They suggested that supportive tools could help foster more natural communication by encouraging appropriate eye contact.

\subsection{Design Study}
\label{sec:3.2}
\par Building upon insights from the semi-structured interviews, we conducted a participatory workshop to translate neurocognitive challenges and needs into assistive design features to address \textbf{RQ2}. The workshop involved 13 participants: 8 adults with ADHD who met the DSM-5 diagnostic criteria (\textbf{P1–P8}), 3 HCI researchers (\textbf{R1–R3}), and 2 clinicians specializing in ADHD (\textbf{C1–C2}). All 8 ADHD participants had previously participated in the semi-structured interviews. Among them, 3 had backgrounds in design, 3 in computer science, and 2 in biology. This interdisciplinary composition was designed to minimize potential biases toward neurotypical interaction norms during ideation.

\subsubsection{Exploratory Brainstorming}

\par To explore a broad range of design ideas for addressing communication challenges related to ADHD, we conducted a group brainstorming session guided by three established principles~\cite{wilson2013brainstorming}: (1) \textit{focus on quantity}, (2) \textit{defer judgment}, and (3) \textit{encourage unconventional thinking}. Participants were encouraged to contribute freely without critique, while also building upon each other's ideas to foster creative synergy. Drawing from the prior interviews, the brainstorming session began with an exploration of commonly reported communication difficulties experienced by individuals with ADHD. Discussions addressed both the limitations of existing interventions and the specific needs for support. Participants were encouraged to generate design concepts for each identified challenge while adhering to the principle of suspending judgment and allowing unrestricted ideation. All ideas proposed during the session were systematically documented. Following the session, researchers and participants collaboratively reviewed and refined the collected concepts, synthesizing a subset of proposals with the greatest potential to address the identified challenges.

\begin{figure}[h]
    \centering
    \includegraphics[width=\linewidth]{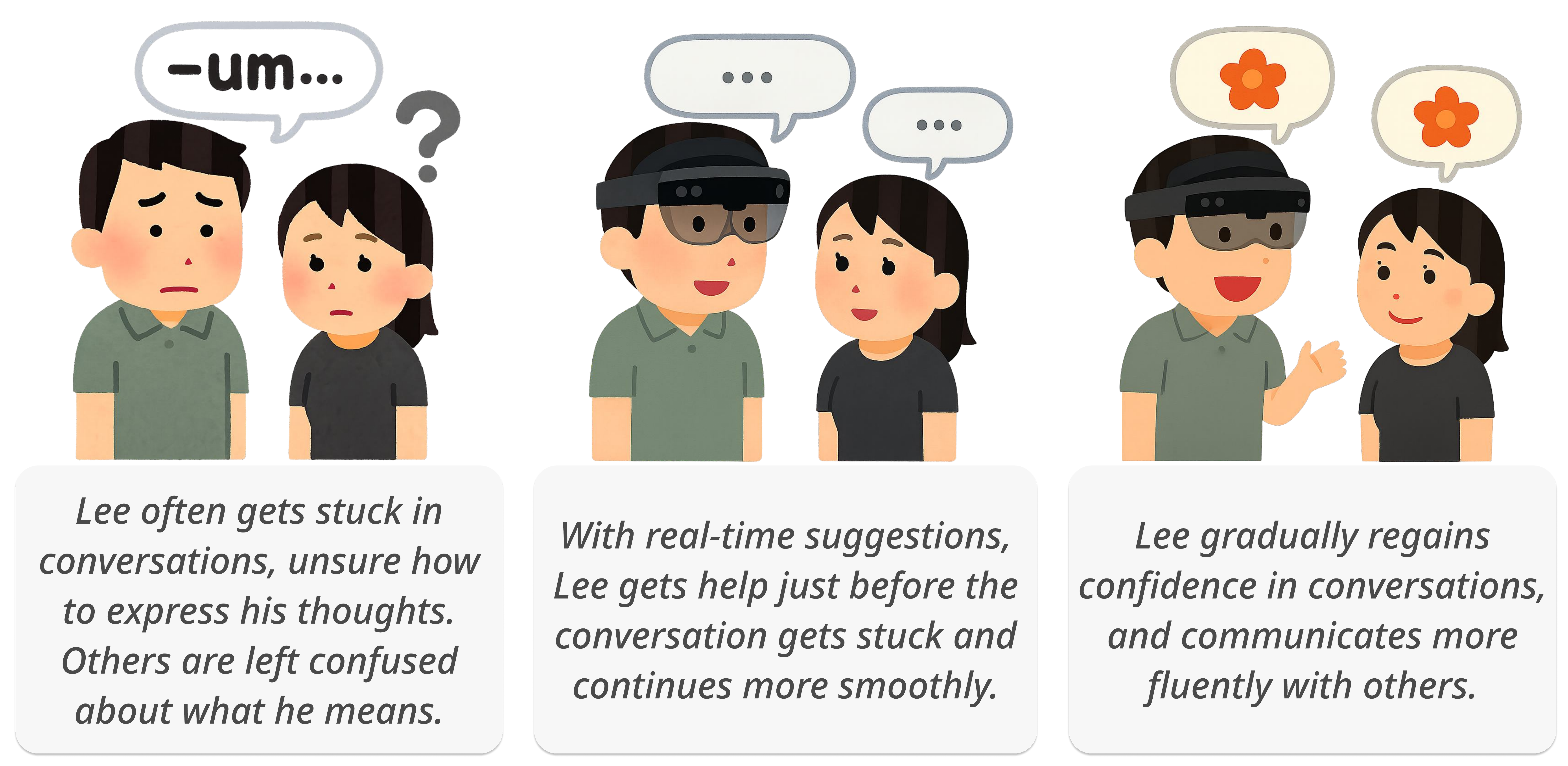}
    \caption{Storyboard No.4: Provides real-time word suggestions based on the context. Participants reported that this feature improved their communication fluency.}
    \label{fig:StoryboardMini}
    \Description{Illustration depicting a storyboard (No. 4) showing the progression of a person's communication with real-time word suggestions. First panel: A person named Lee looks uneasy saying “um…” while another person appears confused with a question mark, indicating Lee’s struggle to express thoughts, leaving others confused. Second panel: Lee, wearing a virtual reality headset, receives real-time suggestions (represented by ellipsis), helping the conversation get unstuck and flow more smoothly with another person. Third panel: Lee, still with the headset, confidently communicates with flower icons above, indicating his confidence and fluency in conversations with others.
    Caption for each panel in the image: (1) Lee often gets stuck in conversations, unsure how to express his thoughts. Others are left confused about what he means. (2) With real-time suggestions, Lee gets help just before the conversation gets stuck and continues more smoothly. (3) Lee gradually regains confidence in conversations, and communicates more fluently with others.}
\end{figure}

\subsubsection{Design Workshop}
\par After formulating initial, high-level design concepts, we refined and expanded these ideas through a dedicated design workshop. Storyboards served as a effective rapid visualization tool~\cite{truong2006storyboarding}, offering a concrete medium to convey and deepen understanding of abstract design concepts. The workshop featured 18 storyboards (examples shown in \autoref{fig:StoryboardMini}, with a summary at \autoref{fig:allStoryboards} in \autoref{app:allStoryboards}), each depicting a communication challenge, a proposed solution, or a hypothetical outcome. These storyboards were iteratively developed from the preliminary design ideas, representing concepts at varying levels of granularity.

\par The workshop began with a review of the communication challenges, and design concepts generated during the earlier brainstorming session. One of the authors shared their screen to present the storyboards in a randomized order. For each storyboard, the presenter introduced the corresponding scenario, outlining the specific communication challenge and the proposed solution. This structured presentation enabled that participants clearly understood the narrative and design intent behind each storyboard.

\par Following each storyboard presentation, participants evaluated the proposed solution, providing insights into its strengths, limitations, and potential applicability. Open discussions were encouraged to elicit constructive feedback on how each concept could be refined or adapted for greater effectiveness.

\par Building on this collective input, participants were then asked to sketch comprehensive design solutions that addressed multiple communication challenges. These designs were iteratively refined based on peer feedback throughout the session. The authors actively documented key insights and unexpected yet valuable contributions from participants, further enriching the understanding of the problem space.

\par Subsequent speed-dating interviews were conducted in accordance with IRB guidelines. Participants received a compensation of \$25 USD for their involvement. All sessions were video-recorded and transcribed for detailed analysis. We systematically organized and categorized the data using affinity diagramming in \textit{Miro}. Regular team meetings were held to interpret findings and reach consensus. Upon completing the analysis, participants were invited to review and validate the results, ensuring alignment with their original perspectives and reinforcing the authenticity and accuracy of the study's conclusions.

\subsubsection{Contextual Analysis Findings}
\par Based on our analysis, we identified several key insights regarding the use of MR devices and strategies for enhancing communication support for individuals with ADHD. Participants expressed a strong preference for MR technologies to deliver real-time assistance, particularly highlighting features such as contextual summarization and word suggestion. However, concerns were raised about managing distractions and ensuring the appropriate delivery of rewards within the system design.

\par \textbf{F3. MR Devices Support Seamless, Polite Interaction and Real-World Skill Training and Transfer.} Participants including \textbf{P1}, \textbf{P3}, and \textbf{R1} considered laptops and smartphones as impractical for daily communication support, as shifting gaze between the device and conversation partner could be socially inappropriate. While teleprompters were conceptually appealing, their usability was limited by spatial constraints in typical communication settings. Regarding contextual support via ear-worn devices like \textit{Memoro}~\cite{zulfikar2024memoro}, participants appreciated the reduced visual salience, which may lessen users' social stigma. However, they expressed concerns about the cognitive demands of processing sequential audio streams. In real-time conversations, managing both the audio streams of partner speech and system prompts was perceived as overwhelming. \textbf{P2} commented, \textit{``As someone with ADHD myself, my brain just ... can't handle that. But combining audio with visual cues might work for me.''} Similar concerns about the cognitive load of auditory versus visual cues have been reported in prior work~\cite{villa2024envisioning,leahy2016cognitive,klingner2011effects}.

\par In contrast, MR devices such as the \textit{HoloLens}\footnote{\url{https://learn.microsoft.com/en-us/hololens/}} were favored for their ability to overlay information directly within the user's field of view, allowing for more natural interactions, such as maintaining eye contact -- an essential aspect of effective communication. As \textbf{P6} remarked, ``\textit{It's visually intuitive to present prompts via MR while minimizing conversational disruption. I can engage with my conversation partner while seamlessly receiving information through MR interfaces -- without extra effort.''} Clinicians also recognized MR's potential to bridge the gap \textbf{G1} in sustainable transfer of therapeutic outcomes. As \textbf{C1} explained, ``\textit{MR enables SST in real-world scenarios, providing opportunities for ADHD people to learn in context and directly apply their skills in everyday interactions.''}

\par \textbf{F4. Contextual Summarization Compensates for Memory-Related Challenges.} In response to \textbf{Ch1}, all participants endorsed the idea of providing conversation summaries to help users recall context and prepare for ongoing dialogue. \textbf{P5} described it as ``\textit{an external memory aid.''} Several participants emphasized that summaries should prioritize recent dialogue while gradually fading earlier content. \textbf{R2} explained, ``\textit{Earlier parts can fade into the background while more recent ones remain visible, as recent ones are more easily forgotten due to attention deficits.''} \textbf{P1} added, ``\textit{Displaying all details would be overwhelming and counterproductive.''}

\par \textbf{F5. Sequential Word Suggestion Mitigate Speech Disfluency.} In addressing \textbf{Ch2}, some participants proposed using visual cues, such as images, to trigger memory recall when speech was blocked. However, most favored a more direct approach: sequential word or phrase suggestions to facilitate sentence completion. \textbf{P4} argued, ``\textit{Instead of processing an image, why not display the word directly?}'' Clinicians echoed this sentiment, with \textbf{C1} noting, ``\textit{When users are overwhelmed and struggle to organize their thoughts, an external hint can help them regain flow.}'' Nonetheless, concerns were raised about potential inaccuracies in word suggestions. \textbf{R3} cautioned, \textit{``If suggestions are off-topic, they might introduce confusion and increase cognitive burden.''}

\par \textbf{F6. Subtle, Non-Intrusive Reminders Help Users Stay on Topic.} With respect to \textbf{Ch3}, most participants recognized the value of detecting topic shifts and offering reminders when users deviated from the conversation. \textbf{R2} suggested audio-based prompts for real-time feedback, but others were concerned about their social appropriateness. \textbf{P3} remarked, ``\textit{A sudden reminder in the middle of a conversation could be disruptive or even uncomfortable.}'' Instead, \textbf{P4} proposed using visual cues as passive, less intrusive reminders. \textbf{R1} built on this idea by suggesting visually encoding the degree of off-topic deviation. \textbf{P4} elaborated, ``\textit{A visual object could serve as a subtle reminder when glimpsed from the corner of the eye. It wouldn't force users to change topics immediately but would allow them to gradually steer back at their own pace.}'' 

\par \textbf{F7. Intuitive Delivery Enhances Comprehension Speed.} Participants unanimously emphasized the importance of rapid response in facilitating effective communication while minimizing cognitive load. Auditory delivery of suggested words was generally disfavored due to slower processing speed and higher cognitive demands. Instead, \textbf{P1}, \textbf{P3}, \textbf{P5}, and \textbf{R3} preferred visual delivery using large fonts and high-contrast colors, describing it as \textit{``visually intuitive and user-friendly''}, allowing users to \textit{``grasp content at a glance''}.

\par \textbf{F8. Minimalistic Design Helps Manage Distractions and Maintain Focus.} Most participants believed that a clean, minimalistic interface would enhance concentration and reduce external distractions. \textbf{P2} suggested that support features should be activated only on demand to avoid cognitive overload, aligning with \textbf{G3} on minimizing system intrusiveness. Additionally, participants recommended automatic deactivation of features after a brief period to conserve cognitive effort and maintain a clutter-free interface. However, \textbf{P6} warned that abrupt disappearance of content could be disorienting, emphasizing the need for controlled transitions to maintain user confidence and a sense of agency.

\par \textbf{F9. Positive Feedback Enhances Confidence and Motivation.} Clinicians (\textbf{C1}, \textbf{C2}) emphasized the therapeutic value of incorporating positive feedback, noting that similar strategies in SST have proven effective in mitigating stigma and frustration stemming from past negative experiences. They suggested applying this principle in system design to foster confidence and encourage engagement. All participants agreed that affirming feedback could motivate users to improve their interpersonal communication skills.

\par \textbf{F10. Customizable Support Features Enhance System Inclusivity.} As highlighted in the semi-structured interviews (\autoref{sec:3.1}), participants exhibited varied communication challenges, leading some to advocate for customizable support features. \textbf{P8} noted, \textit{``I didn't have some of those symptoms, so having support features for them in the system may not be very meaningful to me -- it might even be distracting. That's why I would prefer if users could choose the features they actually need.''} This aligns with \textbf{G2}, which emphasizes empowering user autonomy. Allowing users to tailor support features could better accommodate the evolving nature of ADHD-related communication needs.

\subsection{Design Goals}
\par Drawing from the identified challenges and contextual findings, we derived the following design goals for our communication support system.
\par \textbf{DG1. Accessible, Therapist-Free, Real-Time, and Customizable Support (G1; F1, F2, F3, F10).} The system should provide on-demand, real-time support without requiring therapist intervention. Additionally, it should offer customizable functionalities to accommodate individual user needs. MR devices are particularly well-suited for delivering such adaptive and personalized support.
\par \textbf{DG2. Conversational Content Summarization (Ch1, F4).} To address working memory limitations, the system should provide concise summaries of prior conversations, helping users retain context and engage more effectively in ongoing dialogue.
\par \textbf{DG3. Predictive Word Suggestions for Speech Fluency (Ch2, F5).} To assist users experiencing speech disfluency, the system should offer sequential word or phrase suggestions, facilitating smoother and more confident communication.
\par \textbf{DG4. Off-Topic Detection with Subtle Reminders (Ch3, F6).} The system should identify when conversations deviate from the intended topic and provide non-intrusive, contextually appropriate cues to help users steer discussions back on track.
\par \textbf{DG5. Intuitive, Minimalistic, and Encouraging Interface (G2, G3; F7, F8, F9).} The interface should prioritize intuitive interaction and a minimalistic design to reduce cognitive load while incorporating positive feedback mechanisms to boost user confidence and motivation.

\begin{figure*}[h]
    \centering
    \includegraphics[width=\linewidth]{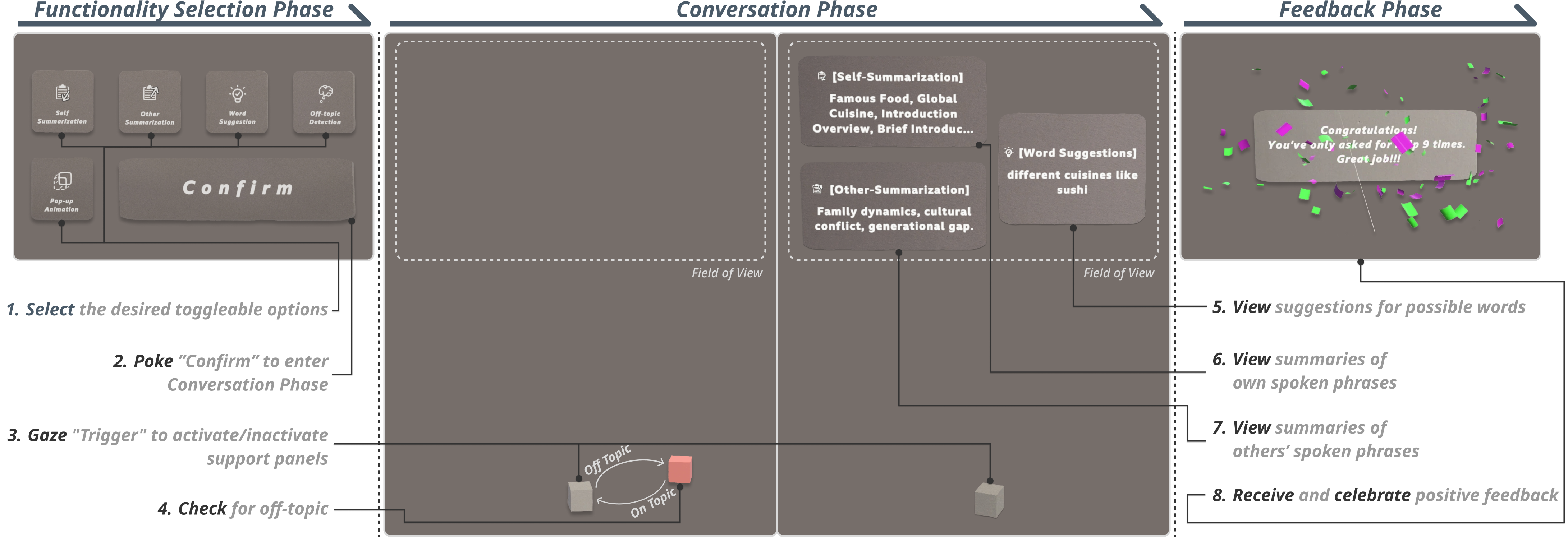}
    \caption{Overview of the full interaction flow in \textit{Understood}. In the Functionality Selection Phase, users can (1) select desired toggleable options and (2) poke ``Confirm'' to proceed. During the Conversation Phase, users (3) gaze at the trigger element to activate or deactivate support panels and (4) check its color to aware of off-topic moments. When panels are activated, users can (5) view word suggestions, (6) view summaries of their own spoken phrases, and (7) view summaries of others' spoken phrases. In the final Feedback Phase, user (8) receive and celebrate positive feedback.}
    \label{fig:SystemOverview}
    \Description{Illustration depicting the interaction flow in Understood, divided into three distinct phases with a clean, structured layout. The Functionality Selection Phase features a horizontal panel with five toggleable icons (Self Summarization, Other Summarization, Word Suggestion, Off-topic Detection) on the left, leading to a prominent “Confirm” button in the center. The Conversation Phase occupies a large central area with a gray field of view, showcasing a trigger element (a small cube) that changes color to indicate on/off-topic status, connected to support panels displaying word suggestions and summaries. The Feedback Phase, on the right, presents a vibrant section with a congratulatory message and colorful confetti icons, all framed within clear, labeled sections and directional arrows guiding the user flow. The caption describes the interaction flow.}
\end{figure*}

\begin{figure}[h]
    \centering
    \includegraphics[width=\linewidth]{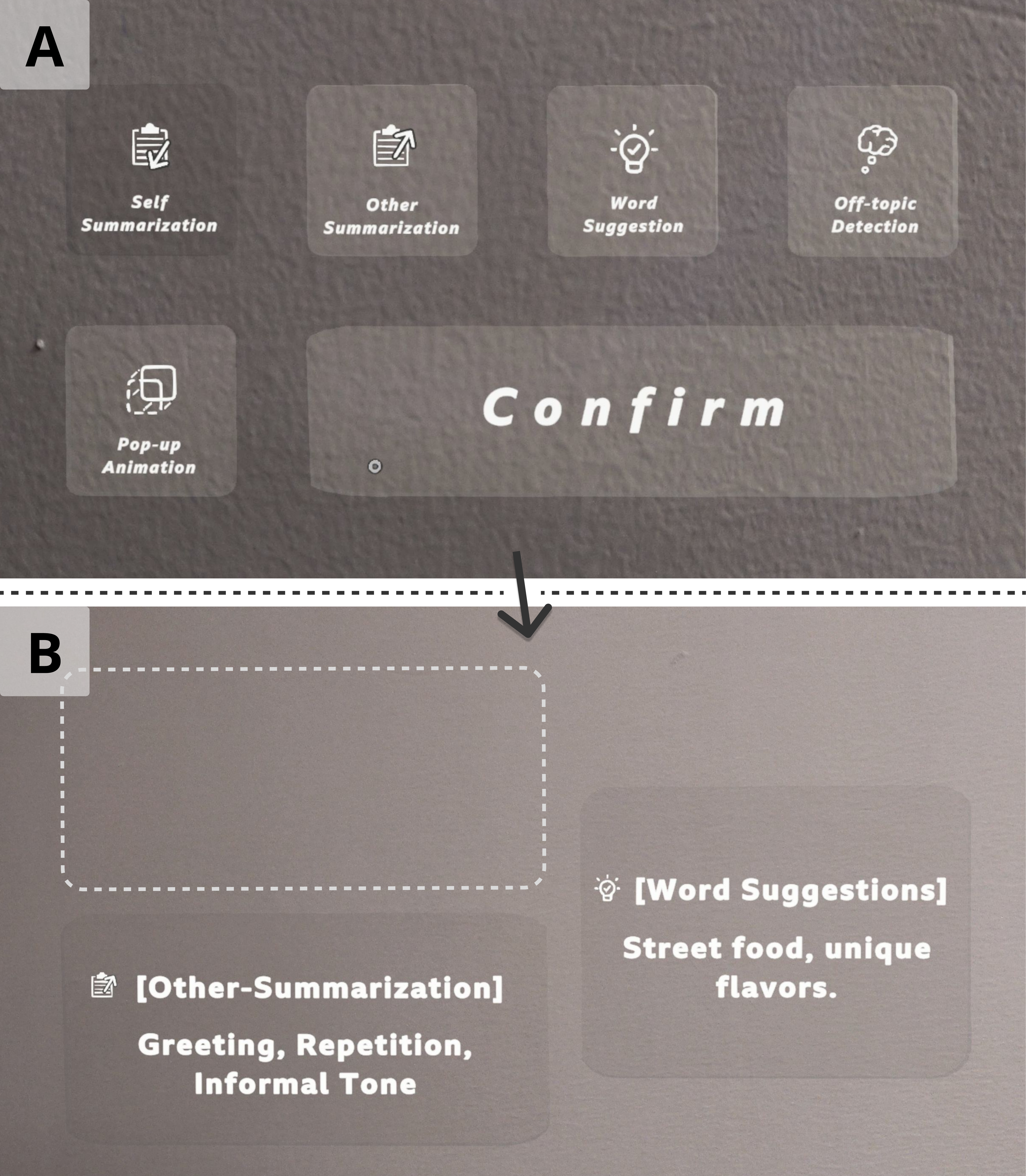}
    \caption{Example of Auxiliary Functions Selection: A) The ``Self-Summarization'' function was deselected during the Functionality Selection Phase. B) The interface in the Conversation Phase after the removal of ``Self-Summarization'' function (where white dashed box indicating the empty space).}
    \label{fig:FunctionSelection}
    \Description{Illustration of the interface for the Auxiliary Functions Selection in Understood, split into two panels labeled A and B. Panel A shows the Functionality Selection Phase with a gray background, featuring five toggleable icons as described in the previous figure. Notably the "Self-Summarization" button was disabled. Panel B depicts the Conversation Phase interface on a gray background, with a white dashed box indicating the empty space where “Self-Summarization” was removed, alongside a tile for “[Other-Summarization]” listing “Greeting, Repetition, Informal Tone” and another tile for “[Word Suggestions]” offering “Street food, unique flavors.”}
\end{figure}

\section{\textit{Understood}}
\subsection{System Overview}
\par Building on insights from the design workshop, we developed \textit{Understood}, a MR system designed to support individuals with ADHD in proactively seeking assistance, enhancing communication fluency, and improving conversational performance, which is available on GitHub~\footnote{https://github.com/Yu-xinz/UIST2025-8878-Understood}. By offering interactive training in realistic settings, the system enables users to gradually refine their speech patterns, serving as a form of automated substitution therapy (\textbf{DG1}). Interaction with \textit{Understood} is structured into three main phases (\autoref{fig:SystemOverview}):
\par \textbf{[1] Functionality Selection Phase.} Upon entering a conversation, users are presented with a customizable set of support features tailored to their individual needs, such as auxiliary tools and motion-based effects. The selected features determine the configuration of the communication support delivered in the subsequent conversation phase (\autoref{sec:Real-timeConversation}).

\par \textbf{[2] Conversation Phase.} Utilizing Azure-based speech recognition, the system continuously monitors the dialogue and dynamically delivers support based on the user's pre-selected features. If users experience difficulty during the conversation, they can trigger assistive prompts via eye-gaze activation, enabling them to access timely, context-aware support and maintain the flow of conversation.

\par \textbf{[3] Feedback Phase.} At the end of the conversation, a feedback window summarizes the user's interactions with the system, including the frequency of assistance requests. This phase offers positive reinforcement, celebrating proactive engagement and reducing the sense of pressure. By framing the experience constructively, the system boosts user confidence for future real-world interactions.

\subsection{Interface and Interaction Design}
\par
\begin{figure}[h]
    \centering
    \includegraphics[width=\linewidth]{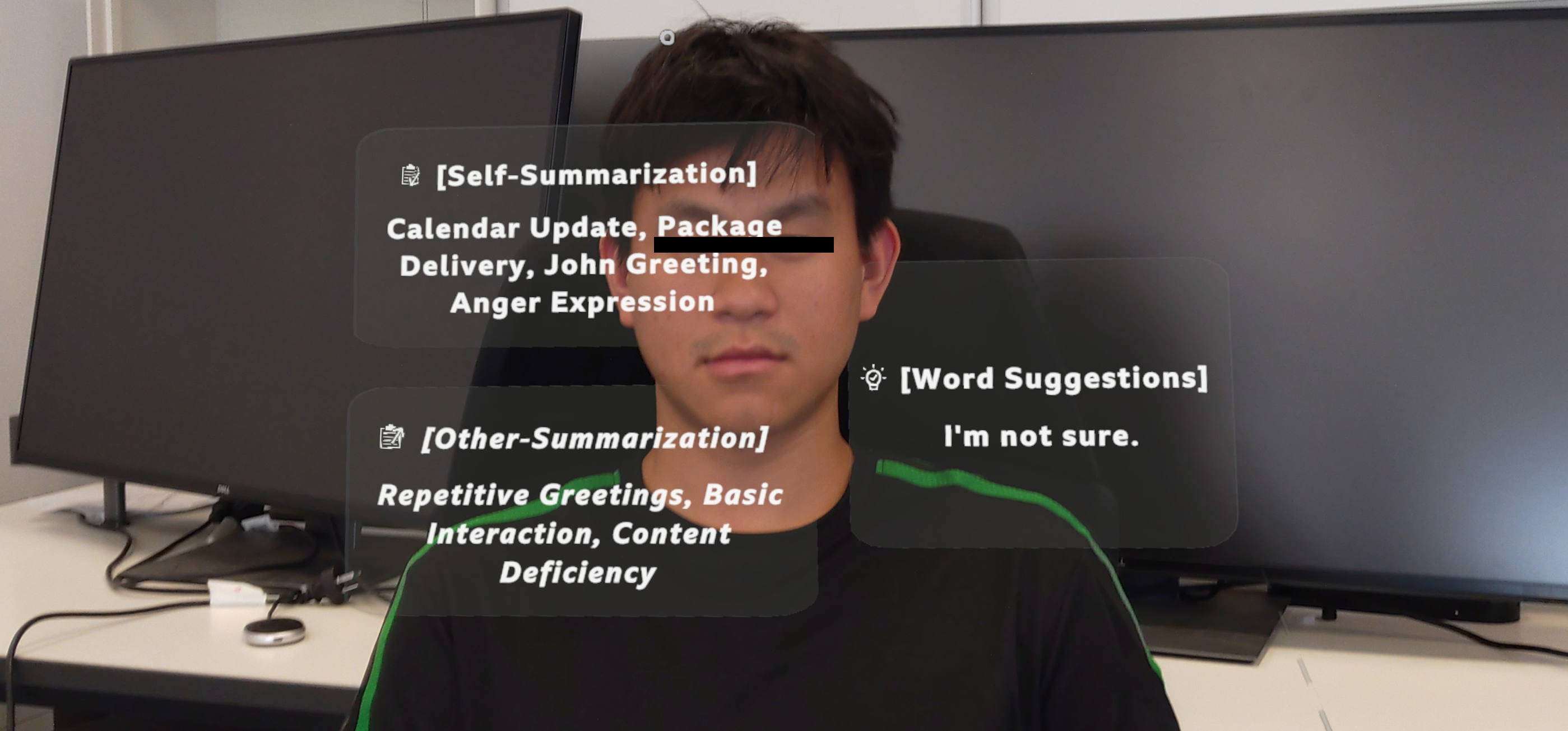}
    \caption{An example usage scenario of \textit{Understood} under a real-world setting.}
    \label{fig:RealisticUsage}
    \Description{Illustration of a real-world usage scenario of Understood, featuring a person sitting at a desk with two computer monitors. The individual is overlaid with translucent interface elements: a “[Self-Summarization]” panel on the left listing “Calendar Update, Package Delivery, John Greeting, Anger Expression,” and an “[Other-Summarization]” panel below noting “Repetitive Greetings, Basic Interaction, Content Deficiency.” A “[Word Suggestions]” panel on the right displays the phrase “I'm not sure.” This demonstrated that Understood's cues could remain clearly visible against a complicated background.}
\end{figure}

 To accommodate the complexity of MR environments -- where real-world backgrounds can introduce distractions, we adopted a minimalist and user-friendly design language. The interface employs a semi-transparent black-and-white color scheme to maintain visual clarity while minimizing cognitive load. In response to potential strain from bright backgrounds, a dark mode was implemented, featuring white text on a black background to reduce visual fatigue. Rounded corners are used consistently throughout the interface to evoke a sense of softness and approachability, ensuring the design remains unobtrusive. To ensure text accessibility, we followed established design guidelines~\cite{williams2020typeface,garnham2017typography,10.1145/3341163.3347748} and evaluated several factors, including design consistency, legibility and readability. Based on these criteria, we selected \textit{BBC Reith}~\cite{bbc_reith} as the system font. Additionally, to mitigate the lack of haptic feedback in MR environments, auditory cues were integrated to provide users with immediate and perceptible confirmation of their actions (\textbf{DG5}). A real-world usage scenario is shown in \autoref{fig:RealisticUsage}. The individual shown provided informed consent for publication.

\begin{table}[h]
\centering
\begin{tabular}{|c|l|l|}
\hline
\textbf{Icon} & \textbf{Function} & \textbf{Description} \\ \hline
\parbox[c]{2em}{\centering\includegraphics[width=2em]{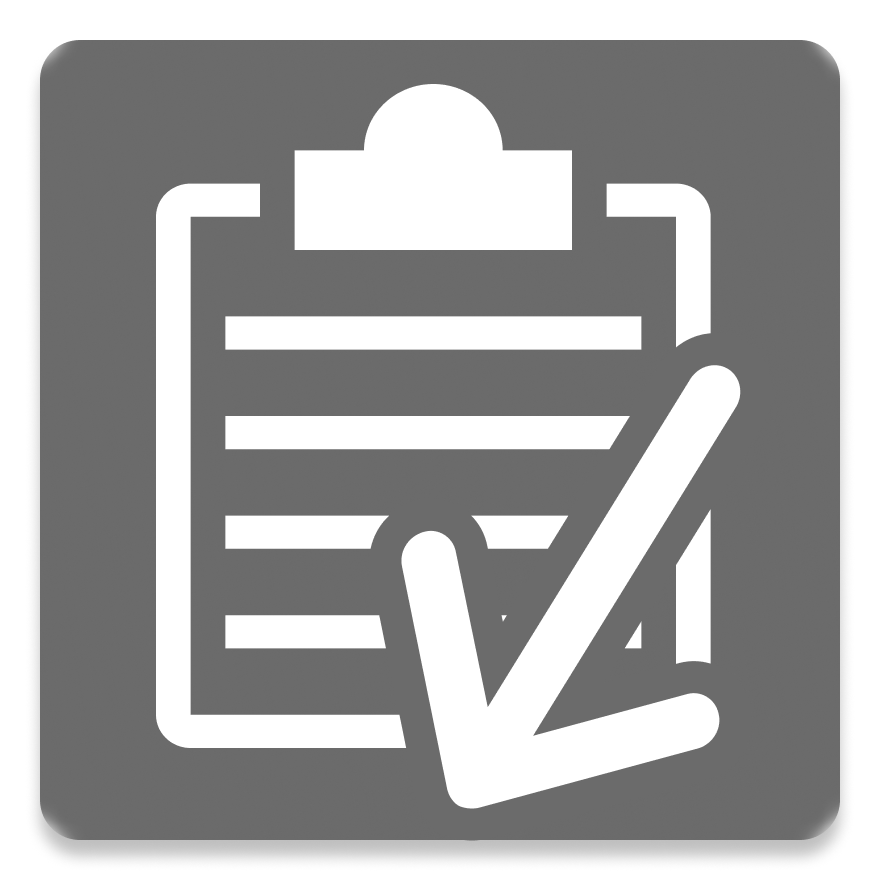}\Description{Icon for Self Summarization.}}  & Self Summarization   & Summarize user's speech                    \\ \hline
\parbox[c]{2em}{\centering\includegraphics[width=2em]{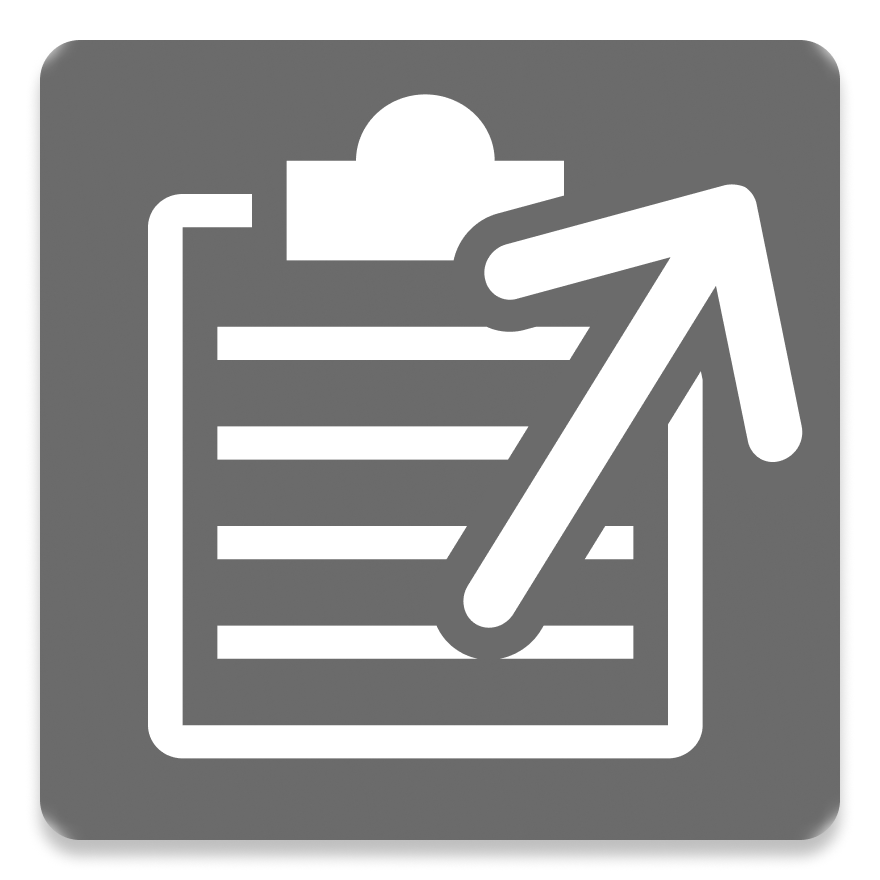}\Description{Icon for Other Summarization.}} & Other Summarization  & Summarize partner's speech                 \\ \hline
\parbox[c]{2em}{\centering\includegraphics[width=2em]{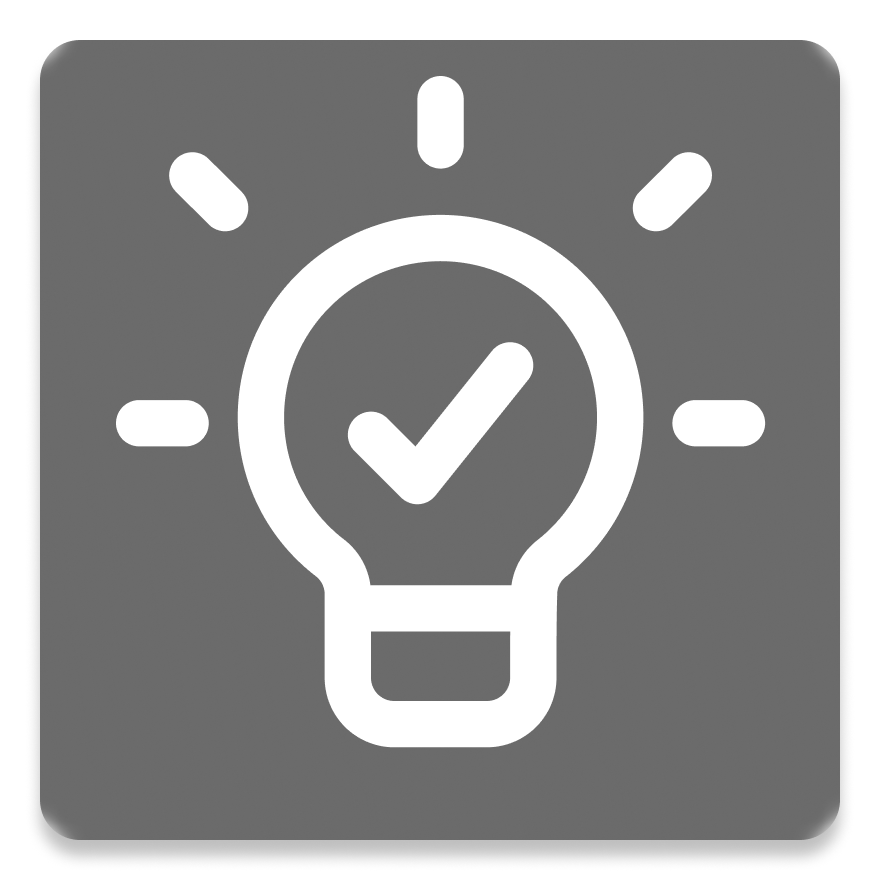}\Description{Icon for Word Suggestions.}}    & Word Suggestions     & Suggest subsequent words             \\ \hline
\parbox[c]{2em}{\centering\includegraphics[width=2em]{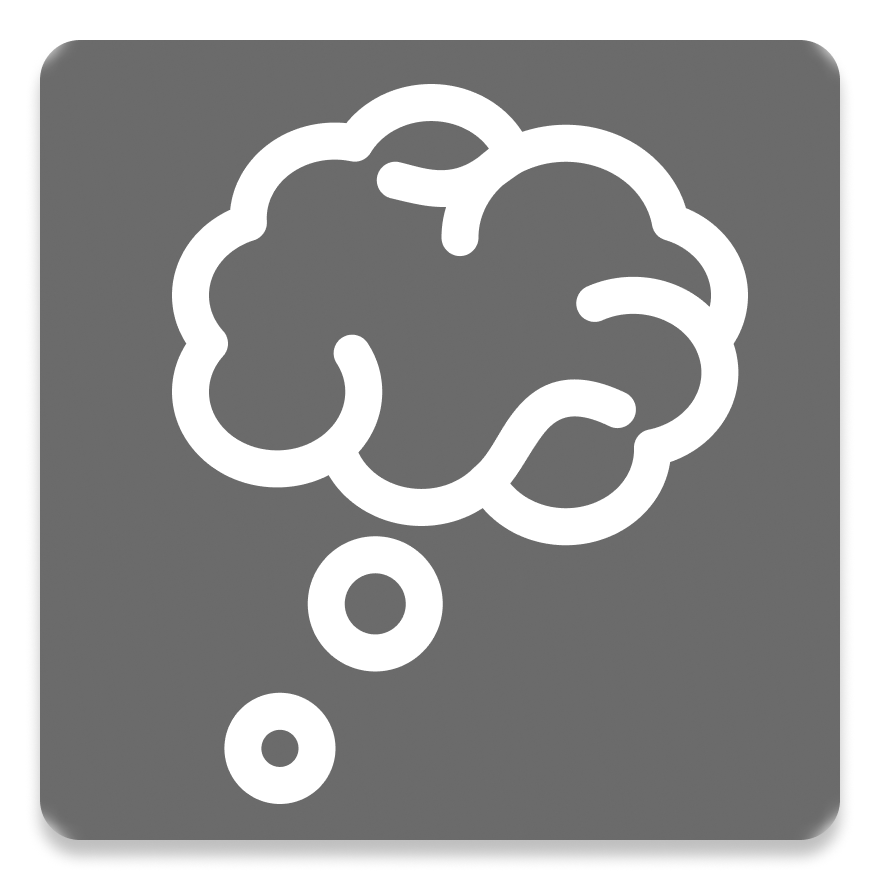}\Description{Icon for Off-topic Detection.}} & Off-topic Detection  & Detect if speech is off-topic           \\ \hline
\parbox[c]{2em}{\centering\includegraphics[width=2em]{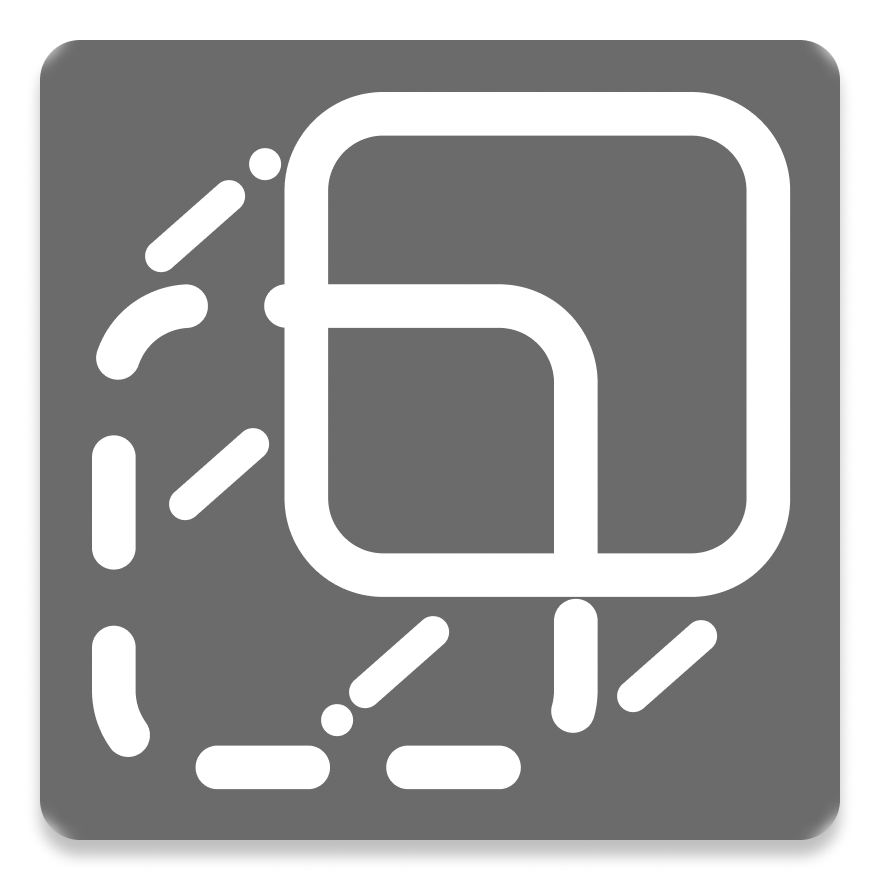}\Description{Icon for Pop-up Animation.}}    & Pop-up Animation     & Pop panel up on gaze             \\ \hline
\end{tabular}
\vspace{0.5em}
\caption{Icon mapping for each auxiliary function.}
\label{tab:IconMapping}
\Description{The mapping between each function, its icon and its description.}
\end{table}
\vspace{-1.9em}

\subsubsection{Auxiliary Functions Selection}
\label{sec:FunctionsSelection}
\par To accommodate users' diverse needs, the auxiliary functions interface offers five toggleable options (\autoref{fig:SystemOverview}-1): four auxiliary tools -- self-summarization, other-summarization, word suggestions, and off-topic detection, as well as one motion effect -- pop-up animation. To enhance intuitiveness, each option is accompanied by a dedicated icon visually representing its function, as shown in \autoref{tab:IconMapping}. All features are enabled by default, allowing users to selectively deactivate any that they find unnecessary or distracting.

\par The interface uses visual opacity to indicate the toggle status of each function: lower opacity denotes that a function is active, while higher opacity signifies that it is deactivated (\autoref{fig:FunctionSelection}). After configuring their preferences, users confirm their selections by poking the ``Confirm'' button (\autoref{fig:SystemOverview}-2), which saves their customized settings for the upcoming conversation phase (\autoref{sec:Real-timeConversation}).

\par This design empowers users to personalize their interaction experience while maintaining a clear overview of active features. By minimizing unnecessary interruptions, the system helps reduce cognitive load and avoids overstimulation, particularly beneficial for users sensitive to excessive sensory input (\textbf{F10, DG1}).

\begin{figure}[h]
    \centering
    \includegraphics[width=\linewidth]{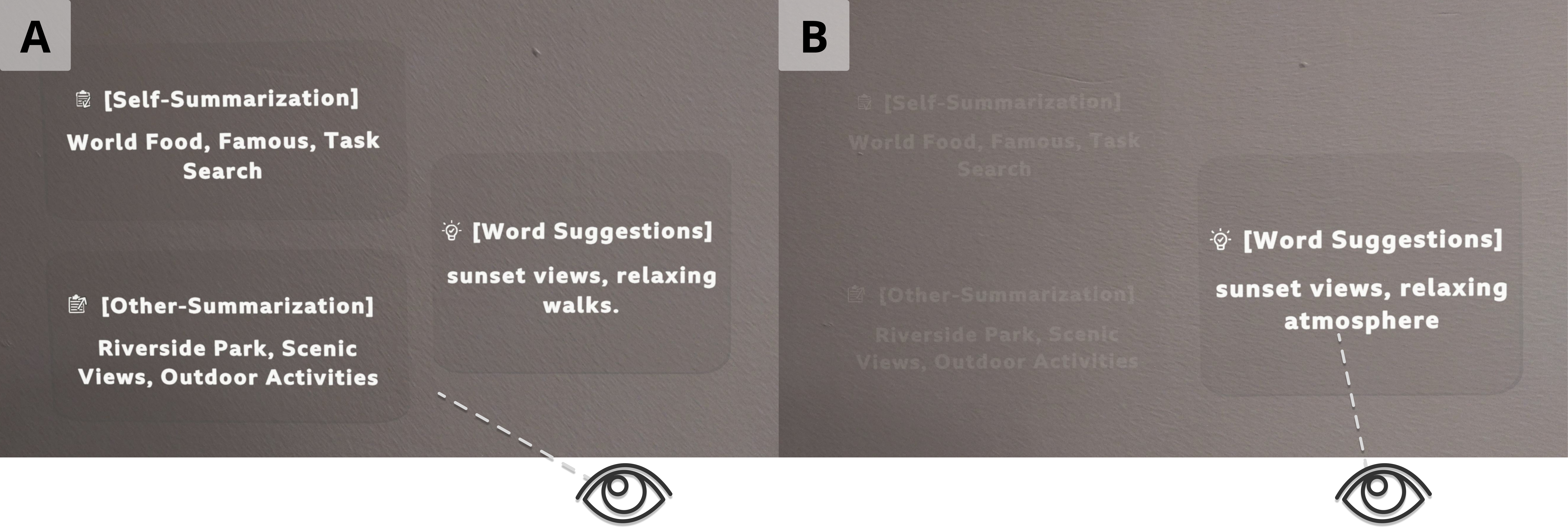}
    \caption{Conversation Phase Interface: A) No gaze directed at any of the panels. B) Gaze directed at the Word Suggestions Panel.}
    \label{fig:Pop-upAnimation}
    \Description{Illustration of the Conversation Phase interface for Understood, split into two sections labeled A and B on a gray background. Section A shows all three panels “[Self-Summarization]”, “[Other-Summarization]” and “[Word Suggestions]” when no gaze directed, indicated by an eye icon away from the panels. Section B mirrors the same summaries but shows a gaze directed at the “[Word Suggestions]” panel, now only highlighting it. The other two un-gazed panels are faded.}
\end{figure}

\subsubsection{Real-time Conversation Support}
\label{sec:Real-timeConversation}
\par Upon entering the interface, a subtle trigger element (\autoref{fig:SystemOverview}-3) appears at the lower part of the user's visual field. This trigger remains minimally intrusive, gently oscillating vertically unless the user intentionally shifts their gaze downward. When assistance is needed during a conversation, users can glance down to activate their pre-selected auxiliary panels, as configured during the Function Selection Phase (\autoref{sec:FunctionsSelection}). These panels then appear within the user's view, providing real-time, context-specific assistance, and automatically fade out after five seconds. Alternatively, users may glance downward again to manually dismiss the panels, thereby maintaining control over the interaction.

\par Once the panels are activated, the trigger element (\autoref{fig:SystemOverview}-3) responds with more dynamic visual behavior: its vertical oscillation increases in amplitude, accompanied by rotation and scale changes, signaling that the system is in an ``active'' state. When multiple auxiliary supports are active, the system also adapts based on eye-gaze behavior. As the user focuses on a particular panel, the remaining panels gradually fade in brightness and increase in transparency, thereby reducing visual distraction (\autoref{fig:Pop-upAnimation}). If the ``Pop-up Animation'' (\autoref{fig:SystemOverview}-1) is enabled, the focused panel will gently pop up, providing additional emphasis. This responsive visual feedback helps users recognize the system's readiness while supporting focused attention on the most relevant aid.

\par Design trade-offs were carefully considered during system development. While placing auxiliary panels in the central field of view may partially obstruct eye contact, this choice was made after exploring and discarding several alternatives during the design workshop(\autoref{sec:3.2}). One proposed approach involved anchoring auxiliary panels beside the interlocutor using body tracking, but implementing real-time body tracking on HoloLens introduced significant system overhead and would likely cause noticeable latency, affecting usability and study validity. Another alternative involved placing panels in peripheral areas of the visual field, but this led to exaggerated gaze shifts during use, which disrupted natural eye contact and made the user appear disengaged in conversation -- an outcome especially undesirable for socially sensitive users.

\par This design reflects two interrelated considerations. First, to minimize distractions during conversations, support panels are not constantly visible -- they appear only when user needed, helping users stay focused on the interaction. Moreover, when multiple panels are active, gaze-based dimming ensures users can concentrate on one feature without overwhelming their field of view (\textbf{F8, DG5}). Second, in face-to-face communication, maintaining appropriate eye contact is crucial for social comfort. By positioning support panels directly in front of the user, the system allows users to seek help discreetly, preserving natural conversational flow while providing meaningful assistance (\textbf{F3, DG5}).

\paragraph{\textbf{Summarization Panel.}} The Summarization Panel includes two components: \textit{self-summarization} (\autoref{fig:SystemOverview}-6) and \textit{other-summarization} (\autoref{fig:SystemOverview}-7), which provide concise recaps of the user's own speech and their conversational partner's speech, respectively. Considering the immediacy of dialogue, the system automatically updates these panels after each utterance. At each update, it first extracts the most recent user utterance and then appends the last available summary from the conversation history. This combined input is submitted to \textit{GPT-4o}~\cite{gpt4o} for subsequent processing (the prompt can be seen at \autoref{app:SummarizationPrompt}). The generated output consists of 4 to 12 keyword-style summary terms displayed in the corresponding panel. These keywords act as memory cues, helping users track the conversational flow and recall key points, especially useful during lapses in attention or external distractions \textbf{(F4, DG2)}.

\paragraph{\textbf{Word Suggestions Panel.}} The Word Suggestions Panel (\autoref{fig:SystemOverview}-5) continuously processes the user's speech. Instead of relying solely on the latest utterance, it references the full dialogue history to inform its suggestions. Based on this comprehensive context, it generates concise lexical suggestions -- typically under 6 words via \textit{GPT-4o} (the prompt is provided in \autoref{app:WordSuggestionsPrompt}). To support real-time interaction without introducing cognitive burden, the system adopts a dynamic refresh strategy: it sends requests every 1 second and updates the panel upon receiving responses. This updates are not paused during user gaze, allowing suggestions to remain timely and contextually relevant. While this may occasionally introduce slight visual distractions, it helps prevent outdated suggestions from persisting and causing greater disruption. This design maintains responsiveness while supporting fluid, meaningful conversation (F5, DG3).

\paragraph{\textbf{Off-topic Detection Panel.}} The Off-topic Detection Panel (\autoref{fig:SystemOverview}-4) is integrated into the same trigger element introduced earlier, streamlining visual design and preserving conversation flow. Rather than introducing a separate notification mechanism, which could be disruptive -- particularly for users sensitive to interruptions. Our system continuously monitors ongoing speech and context. It sends the combined input to \textit{GPT-4o} to assess topic relevance (the prompt is given in \autoref{app:Off-topicDetectionPrompt}). If off-topic content is detected, the trigger gradually shifts to a deeper red hue, serving as a soft visual alert. This gentle transition prompts users to self-correct without breaking engagement. By combining help-seeking and topic-awareness into a single interactive element, the design ensures seamless and discreet feedback, encouraging sustained focus and interaction quality \textbf{(F6, DG4)}.

\subsubsection{Status Feedback}
\par Upon completion of the conversation, after poking the trigger element (\autoref{fig:SystemOverview}-3), a pop-up window (\autoref{fig:SystemOverview}-8) appears in the user's field of view, summarizing the number of times assistance was actively sought during the interaction. To deliver immediate positive reinforcement, a celebratory confetti animation is triggered. Each tap on the window generates a new burst of confetti, introducing a sense of interactivity and playfulness. This gamified feedback is designed to lighten the overall experience, reframing the interaction as enjoyable and affirming rather than resembling a clinical training session.

\begin{figure*}
    \centering
    \includegraphics[width=0.8\linewidth]{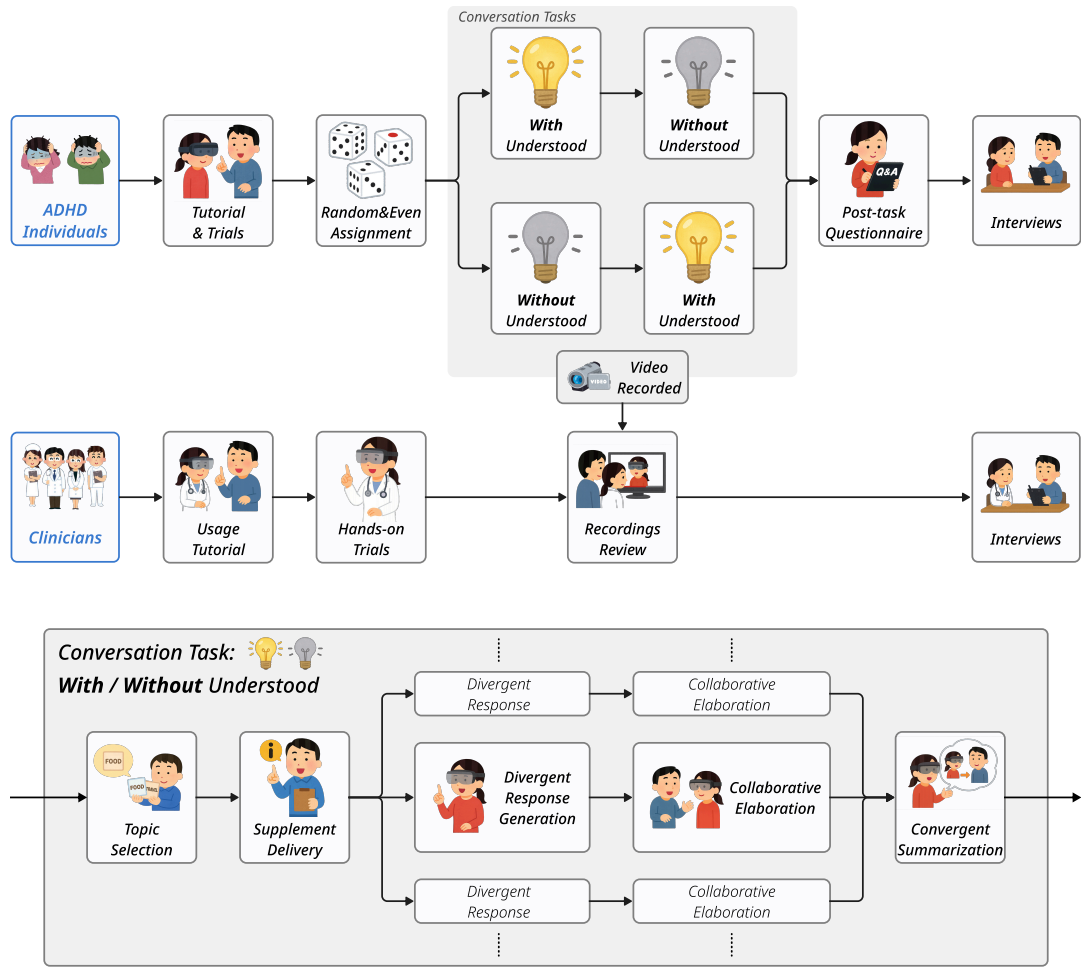}
    \caption{User study and procedure for evaluating the effectiveness and usability of \textit{Understood}. For clarity, the procedures for conversation task (with and without \textit{Understood}) are presented separately at the bottom. The study involved both individuals with ADHD and clinicians, where individuals with ADHD completed a tutorial, two conversation tasks, questionnaires, and interviews, while clinicians reviewed video recordings and provided evaluations.}
    \label{fig:UserStudyProcedure}
    \Description{Illustration of the user study and procedure for evaluating Understood's effectiveness and usability, divided into two main sections. The top section outlines the conversation tasks: ADHD individuals and clinicians start with usage tutorials and trials, followed by random/even assignment to “With Understood” or “Without Understood” groups, depicted with light bulbs. Both groups proceed to post-task questionnaires and interviews. The bottom section details the conversation task process (with/without Understood): starting with topic selection and supplement delivery, followed by divergent response generation and collaborative elaboration, leading to convergent summarization, illustrated with icons and flow arrows. Clinicians' path includes hands-on trials, video recorded recordings review, and interviews.}
\end{figure*}

\par Beyond its playful surface, this design also serves a deeper function. By cultivating a relaxed and non-judgmental atmosphere, the system encourages users to engage in conversations with a positive mindset. This subtle and supportive approach promotes the development of healthy communication habits without explicitly framing the experience as training. In doing so, it helps reduce user reliance on the system and supports the seamless transfer of conversational skills to real-world scenarios where assistive tools may not be available \textbf{(F9, DG5)}.

\subsection{Technical Details}
\par \textit{Understood} is implemented on the \textit{Microsoft HoloLens 2} using the Mixed Reality Toolkit 3~\footnote{https://learn.microsoft.com/en-us/windows/mixed-reality/mrtk-unity/mrtk3-overview} (MRTK3) and Unity (version 2022.3.53f1c1). The system enables interaction through both hand gestures and gaze tracking. For voice input, it leverages Azure AI Speech~\footnote{https://azure.microsoft.com/en-us/products/ai-services/ai-speech} to process spoken dialogue. To support summarization, word suggestion, and off-topic detection, the system integrates \textit{GPT-4o} via \textit{OpenAI}'s API~\footnote{https://platform.openai.com/docs/api-reference/chat}. During the feedback phase, celebratory effects are rendered using the \textit{Confetti FX library}~\footnote{https://assetstore.unity.com/packages/vfx/particles/confetti-fx-82497} from the Unity Asset Store.

\par All spoken dialogue is processed in real time and is neither stored locally nor retained post-session. The audio data is transmitted solely to Azure and OpenAI services, in accordance with the privacy polices of the respective providers.

\section{User Study}
\par To evaluate the \textbf{effectiveness (RQ3)} and \textbf{usability (RQ4)} of \textit{Understood}, we conducted a within-subjects study in which participants completed communication tasks under two conditions: (1) with the support of \textit{Understood}, and (2) without it. This experimental design enabled direct comparison of participant performance across both scenarios.

\subsection{Participants}
\par Following IRB approval, we recruited 10 \underline{individuals diagnosed with} \underline{ADHD} (\textbf{P1-P10}; 4 male, 6 female; mean age = 21.3, SD = 2.24) from a local university community. All participants self-reported experiencing communication challenges corresponding to at least one of the three types (\textbf{Ch1-Ch3}) outlined in our framework. Upon completion of the study, we invited two \underline{clinicians} -- \textbf{C1}, a psychiatrist, and \textbf{C2}, a therapist, both of whom had participated in the formative study -- to use \textit{Understood} and review participant session recordings. Their expert feedback provided a complementary clinical perspective on the system's performance and applicability.

\subsection{Procedure}
\par Informed by prior research~\cite{EVENSIMKIN202466,barkley2007adhd-ch13,kofler2019working,zulfikar2024memoro,wu2024trinity} and developed in collaboration with clinicians \textbf{C1} and \textbf{C2}, we designed the user study procedure illustrated in \autoref{fig:UserStudyProcedure}. We began by obtaining informed consent and collecting basic demographic information, with a focus on participants' communication-related challenges. Each participant was then introduced to the \textit{Understood} system and received a guided walkthrough of its key functionalities while wearing the \textit{HoloLens} device. Following the tutorial, participants were given 10 minutes to freely explore the system and engage in informal conversation with the researchers, providing them an opportunity to familiarize themselves with the interface and features in a relaxed setting.

\subsubsection{Conversation Task Design}
\par Once participants were familiar with the system, they engaged in two open-ended conversation tasks: one supported by \textit{Understood} and one conducted without it. Each conversation lasted approximately 7 minutes and was separated by a short break. To mitigate potential order effects, the study employed a counterbalanced design: the 10 participants were evenly split into two groups -- one began with the \textit{Understood}-assisted conversation, while the other started without system support. All conversations were conducted in participants' native language to avoid confounding influences from foreign language proficiency.

\par Conversation topics were randomly selected from a predefined pool of five, with each session using a different topic. Participants could skip or switch topics if they felt uncomfortable. All topics were designed to be grounded in everyday experiences (e.g., ``\textit{What is your favorite place in the city you currently live in?}''), ensuring participants had sufficient contextual knowledge to engage meaningfully.

\par To simulate challenges related to \textbf{working memory deficits} (\textbf{Ch1}), each topic was accompanied by supplementary information (e.g., ``\textit{As for me, my favorite place is Riverside Park, because ...}''), presented verbally in a 45-second segment. Participants were instructed to retain this information throughout the conversation. A full list of topics and their corresponding supplements is provided in \autoref{app:userstudy-topic}.

\par To replicate the effects of \textbf{attentional anchoring failures} (\textbf{Ch3}), each conversation followed a \textit{divergent-to-convergent} structure. In the divergent phase, participants iteratively generated and elaborated on multiple ideas, with the researcher actively responding to each one. This interactive format naturally encouraged topic shifts and deeper engagement.

\par In the subsequent convergent phase, participants were tasked to summarize the conversation, including both their own input and that of the the researcher. This task increased cognitive loads by introducing multiple subtopics while requiring synthesis and memory recall. Throughout both phases, \textbf{speech disfluencies and pauses} (\textbf{Ch2}) were allowed to arise naturally during conversation.

\subsubsection{Post-Task Measures and Clinician Interview}
\par Upon completing both conversation tasks, participants were asked to fill out a post-task questionnaire evaluating their experience in terms of efficiency and usability. Responses were recorded using a 5-point Likert scale. The questionnaire items were adapted from well-established instruments, including the System Usability Scale (SUS)~\cite{brooke1996sus}, the NASA Task Load Index (NASA-TLX)~\cite{hart1986nasa}, and relevant recent studies~\cite{zulfikar2024memoro,wu2024trinity,cai2023paraglassmenu}. All the question statements in the post-task questionnaire are listed at \autoref{app:userstudy-question}. In addition, semi-structured interviews were conducted to gather qualitative feedback and suggestions. Each participant received a compensation of \$15 USD for their participation.

\par All study sessions were conducted in person and recorded -- with participant consent -- for subsequent analysis. Following the completion of all sessions, two clinicians were invited to engage with the \textit{Understood} system and review recordings of participant interactions. To complement participant feedback with expert insights, we conducted semi-structured interviews with both clinicians. These interviews explored participants' conversational behaviors, emergent usage patterns, and the clinicians' overall impressions of \textit{Understood}. Each clinician was compensated \$40 USD for their time.

\begin{figure*}[h]
    \centering
    \includegraphics[width=\linewidth]{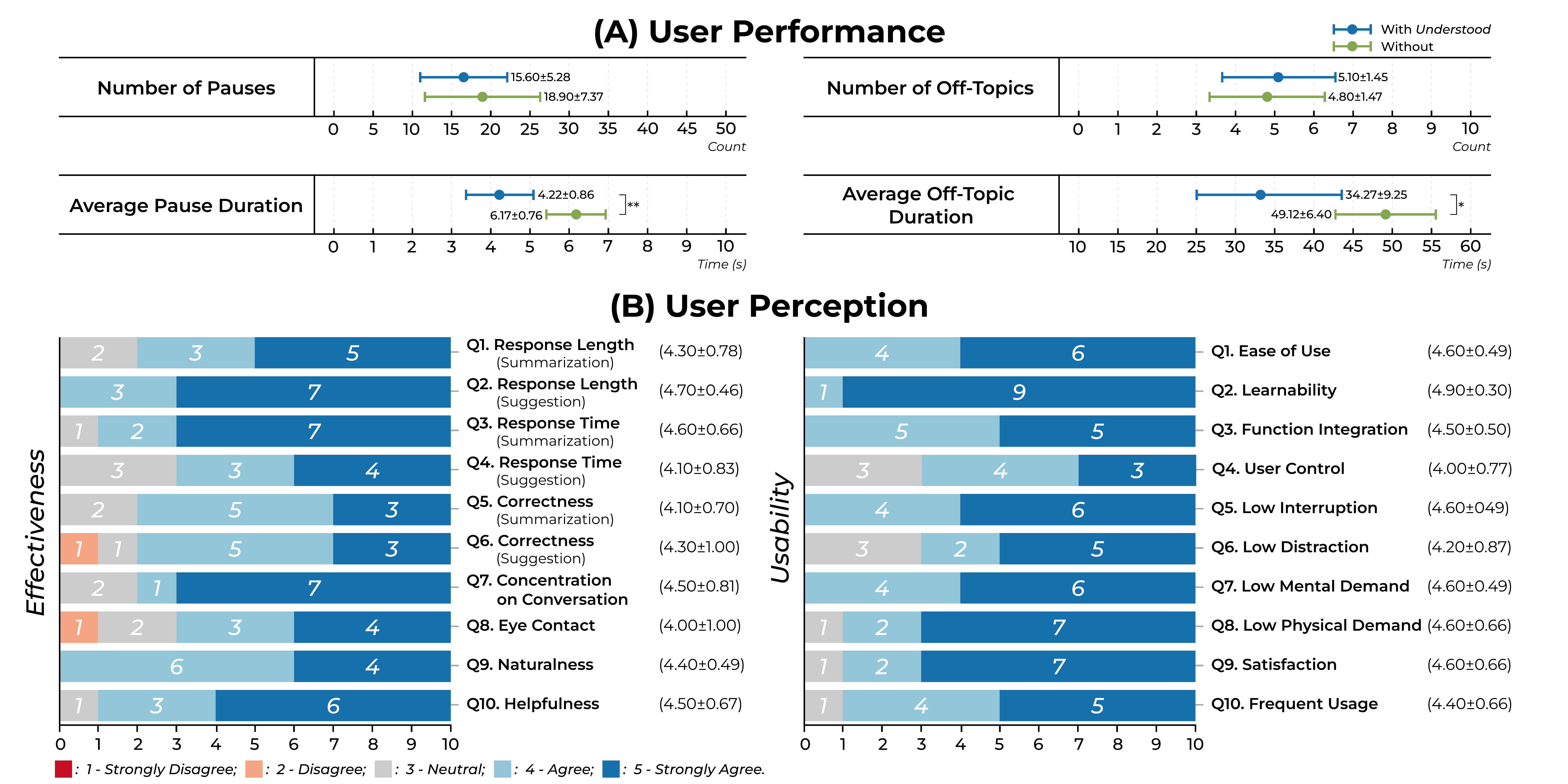}
    \caption{User study results of (A) User Performance with \textit{Understood}, and (B) User Perceptions towards \textit{Understood}. Results indicate that \textit{Understood} effectively support the communication with high usability. ($*:p<0.05;$ $**: p<0.01.$)}
    \Description{Please read the following "Result and Analysis" section for the results reported in the image.}
    \label{fig:UserStudyResult}
\end{figure*}

\subsection{Results and Analysis}

\par In addition to post-task questionnaire ratings, we collected quantitative data to evaluate the performance of both the \textit{Understood} system and its users. Key metrics included: (1) Average LLM Response Length, (2) Average LLM Response Time, (3) Number of Pauses, (4) Average Pause Duration, (5) Number of Off-Topic Responses, and (6) Average Off-Topic Duration. For metrics measured under both conditions (with and without \textit{Understood}), we applied the Wilcoxon matched-pairs signed ranks test to assess whether significant differences existed across these variables (3)-(6).
\par Based on the results presented in \autoref{fig:UserStudyResult} and \autoref{tab:LLMStatistics}, along with insights from semi-structured interviews with both individuals with ADHD and clinicians, we derived the following conclusions.

\begin{table}[h] 
\caption{Average response length (in words) and response time (in seconds) of \textit{Understood} for summarization and suggestion tasks.}
\begin{tabular}{cc|cc}
\multicolumn{2}{c|}{\textbf{Response Length (words)}} & \multicolumn{2}{c}{\textbf{Response Time (s)}} \\ \hline
\textit{Summarization}      & \textit{Suggestion}     & \textit{Summarization}  & \textit{Suggestion}  \\
$8.14\pm1.73$               & $3.52\pm0.88$           & $1.25\pm0.40$           & $1.95\pm0.71$       
\end{tabular}
\label{tab:LLMStatistics}
\end{table}

\subsubsection{Effectiveness}
\par In terms of \textbf{Response Length}, \textit{Understood} generates concise and intuitive prompts (\textit{Summarization}: $8.14\pm1.73$ words; \textit{Suggestion}: $3.52\pm0.88$ words), which were deemed appropriate by users, as reflected in their perception ratings (\textit{Summarization}: $4.30\pm0.78$; \textit{Suggestion}: $4.70\pm0.46$). As \textbf{P5} observed, \textit{``The summarization is brief, never exceeding three lines, and the suggestions are concise, requiring only a few words. I can quickly understand the content and use it to steer my next response without spending too much time reading.''}

\par In terms of \textbf{Response Time}, \textit{Understood} achieved average response times of $1.25\pm0.40$ seconds for \textit{Summarization} and $1.95\pm0.71$ seconds for \textit{Suggestion}. The longer duration for Suggestion is attributed to the reliance on the full conversation history, which increases the processing load as the conversation progresses, as detailed in \autoref{sec:Real-timeConversation}. In contrast, Summarization builds upon previous summaries, keeping the processing time relatively constant. Overall, participants responded positively to the system's responsiveness (ratings: \textit{Summarization}: $4.60\pm0.66$; \textit{Suggestion}: $4.10\pm0.83$). As \textbf{P3} commented, \textit{``Understood responds in a reasonable time frame. Anyone familiar with LLMs would find the timing quite acceptable.''}

\par Regarding \textbf{Correctness}, responses generated by \textit{Understood} were generally perceived as helpful by most participants (\textit{Summarization}: $4.10 \pm 0.70$; \textit{Suggestion}: $4.30 \pm 1.00$). As \textbf{P1} remarked, \textit{``When my mind went blank, Understood often provided a reasonable hint to help me continue. The Summarization helped me organize the context, while the Suggestion provided a new opportunity.''} Several participants also reported that the the responses encouraged them to expand the scope of the conversation. For example, \textit{P4} shared, \textit{``I was asked to talk about my favorite place, but I couldn't immediately think of anything. Then the system suggested `City Park', which reminded me of a wonderful picnic experience from a few years ago. Without the suggestion, I wouldn't have recalled the memory.''}

\par However, due to the inherent hallucination issues in LLMs, the system may occasionally produce irrelevant responses, particularly in situations where the preceding conversation ends with an open-ended question with allows for multiple valid answers. In such cases, the generated suggestions may not align with users' intended direction. As \textbf{P6} noted, \textit{``Sometimes I have a term in mind but struggle to articulate it. At those moments, the system offered suggestions that were unrelated, which left me confused and overwhelmed.''}

\par Additionally, since the suggestion responses were generally promising, some participants were observed to frequently rely on the suggestion function. This raised concerns among clinicians about the risk of over-reliance. As \textbf{C2} cautioned, \textit{``If they constantly use the suggestions instead of thinking for themselves, this could be a troubling sign: they may become dependent on the suggestions, which could lead to a decline in their communication skills.''} To mitigate this risk, auxiliary functions were intentionally positioned outside the user's immediate line of sight, making them less prominent and more difficult to trigger unintentionally. This design choice aimed to discourage excessive reliance on the system. We further elaborate on this issue in \autoref{sec:Discussion-Reliance}.

\par Regarding \textbf{User Performance}, the results suggest that participants benefited from the support of \textit{Understood}. While no statistically significant differences were observed in the \underline{Number of} \underline{Pauses} (With: $15.60 \pm 5.28$; Without: $18.90 \pm 7.37$) and \underline{Number of} \underline{Off-Topics} (With: $5.10 \pm 1.45$; Without: $4.80 \pm 1.47$), participants showed significant improvements in their ability to recover from these events. Specifically, there was a notable reduction in \underline{Average} \underline{Pause Duration} (With: $4.22 \pm 0.86$ seconds; Without: $6.17 \pm 0.76$ seconds; $p = 0.002$) and \underline{Average Off-Topic Duration} (With: $34.27 \pm 9.25$ seconds; Without: $49.12 \pm 6.40$ seconds; $p = 0.014$). These findings suggest that while \textit{Understood} may not significantly alter the frequency of pauses or off-topic responses, it effectively facilitates quicker recovery when such events occur.

\par Clinicians, however, raised concerns about \textit{Understood}'s focus on symptom management rather than addressing the underlying causes of ADHD-related communication challenges. As \textbf{C1} noted, \textit{``While I appreciate Understood's ability to help ADHD people manage communication difficulties, I worry that it only addresses the `surface-level' symptoms of ADHD. A symptom-focused approach may have limited long-term impact. Nonetheless, I see potential in how the system could help ADHD people transfer skills from controlled environments to real-world interactions.''} This concern is further discussed in \autoref{sec:Discussion-Balance}.

\par Participants responded positively to \textit{Understood} in terms of enhancing their \textbf{Concentration on Conversation} ($4.50 \pm 0.81$), \textbf{Eye Contact} ($4.00 \pm 1.00$), and overall \textbf{Naturalness} ($4.40 \pm 0.49$). These results indicate that the system effectively supported participants in maintaining conversational focus and engaging more naturally. \textbf{P9} commented, \textit{``With the summarization feature, I no longer worry about forgetting the previous parts of the conversation, which helps me feel more confident and stay focused while talking.''} Similarly, \textbf{P4} noted, \textit{``I can seamlessly integrate the summarization and suggestions into my speech, and the words naturally flow out of my mouth.''} \textbf{P7} added, \textit{``When my gaze landed on the red-colored trigger, it subtly reminded me that I had gone off-topic, helping me steer the conversation back on track smoothly.''}

\par However, some participants expressed concerns about maintaining eye contact, noting that their gaze often shifted back and forth between the panels. Despite this, clinicians viewed eye contact more positively from an audience perspective. \textbf{C2} remarked, \textit{``From my perspective, I seldom notice their gaze drifting away from the conversational partner, so the effect is quite acceptable.''} This gaze-shifting behavior may also be interpreted as a user-initiated coping strategy. By consciously shifting their gaze between the panels, participants appeared to regulate their attention and recover from conversational breakdowns more effectively.

\par Overall, most participants found \textit{Understood} to be effective in helping them overcome communication challenges, as reflected in the \textbf{Helpfulness} rating of $4.50 \pm 0.67$.

\subsubsection{Usability}

\par In terms of \textbf{Ease of Use} ($4.60 \pm 0.49$) and \textbf{Learnability} ($4.90 \pm 0.30$), all participants agreed that \textit{Understood} provided an intuitive interaction mechanism that helped them overcome communication challenges. The minimalistic design further facilitated quick mastery of the system. \textbf{P4} commented, \textit{``A clever solution! \textit{Understood} tackles challenges with simple, well-defined functions. I was able to grasp its features and usage in a short time, which will be especially helpful for first-time users.''} This feedback is supported by the \textbf{Function Integration} rating of $4.50 \pm 0.50$. Additionally, participants found the \textit{Function Selection Phase} useful, as it allowed them to customize the system’s functions and interface to better meet their individual needs.

\par Regarding \textbf{User Control} ($4.00 \pm 0.77$), most participants felt that \textit{Understood} performed as expected. However, some noted that hallucinations by the LLMs could lead to unexpected responses, causing confusion. \textbf{P6} added, \textit{``When I encounter unexpected responses, I have to put more mental effort into processing them, as I need to filter out these responses from my original thoughts.''} As a result, although \textit{Understood} received a high rating for \textbf{Low Interruption} ($4.60 \pm 0.49$) due to its minimal and well-defined design, the rating for \textbf{Low Distraction} was lower than other statements ($4.20 \pm 0.87$), primarily due to inaccuracies.

\par \textit{Understood} also gained consensus among participants for its low mental ($4.60 \pm 0.49$) and physical ($4.60 \pm 0.66$) demands. \textbf{P3} remarked, \textit{``The summarization and suggestions were incredibly helpful. I no longer need to worry about my personal deficits, as \textit{Understood} provides support. I can finally relax.''} \textbf{P10} mentioned that triggering auxiliary functions by gaze was physically relaxing: \textit{``If you've ever interacted with an AR/VR system, you'll understand how tiring it can be to raise your arms and hands in midair.''}

\par \textbf{Satisfaction} was also highly rated by participants, with a score of $4.60 \pm 0.66$. They found the final \textit{Feedback Phase} both helpful and uplifting. \textbf{P5} noted, \textit{``As the panel displayed the number of times I sought assistance during the conversation, I could use it to evaluate my performance and track my progress. Seeing the number decrease was quite a rewarding experience.''} \textbf{P7} added, \textit{``The confetti animation is really fun! It provides further relaxation after a stressful interpersonal conversation.''} Overall, most participants expressed their willingness for \textbf{Frequent Use} of \textit{Understood} ($4.40 \pm 0.66$).

\subsubsection{Suggestions for Understood}
\par In the semi-structured interviews, participants identified areas for improvement in \textit{Understood}, aiming to enhance its efficiency, user-friendliness, and accessibility.

\paragraph{\textbf{Performance Optimization}} Currently, limitations in LLMs lead to occasional hallucinations, where responses may fail to align with user expectations. While integrating more sophisticated models could enhance accuracy, this may also result in longer response times, potentially disrupting the conversational flow. Participants also suggested improvements in information delivery. For instance, \textbf{P4} recommended offering more structured summaries to clarify the connections between key concepts, while \textbf{P5} proposed adding visual cues alongside textual hints to encourage exploration of new topics. These enhancements could improve \textit{Understood}'s effectiveness, particularly in helping users engage more efficiently.

\paragraph{\textbf{Better Customization}} \textit{Understood} currently enables users to select functionalities tailored to their specific challenges, offering a certain level of customization. However, participants identified potential areas for improvement. \textbf{P4} suggested that \textit{Understood} could offer longer summaries and allow users to scroll through previous conversations for more thorough reviews, which would be especially beneficial for those wanting to revisit specific points. Additionally, \textbf{C2} recommended that \textit{Understood} enhance its ability to transfer CBT/SST outcomes to real-life scenarios by automatically identifying a user's current challenges and adjusting support accordingly. This targeted customization could help individuals apply the strategies they've learned more effectively.

\paragraph{\textbf{Beyond HoloLens}} While \textit{HoloLens} has proven effective in assisting users with communication challenges, it does have certain limitations. Some participants noted that the device's weight could cause discomfort, with \textbf{P8} mentioning that wearing it for extended periods might be difficult. Additionally, the high cost of \textit{HoloLens} could restrict its accessibility to a broader user base. In light of these concerns, participants suggested exploring alternative, more affordable MR platforms to enhance the accessibility of \textit{Understood} and expand its reach to a wider audience.

\section{Discussion}
\subsection{Support vs. Self-Reliance: A Design Tension}
\label{sec:Discussion-Reliance}
\par In our user study, \textit{Understood} provided auxiliary features, such as summarization and suggestions, which were particularly helpful for users during moments of cognitive struggle. However, clinicians raised concerns about the potential for over-reliance on these features, which could hinder the development of self-reliance. For example, the suggestion function offers word candidates when users get stuck, reducing cognitive load and maintaining conversational flow. However, some participants were observed to rely on these suggestions without engaging in independent thought, effectively reducing cognitive involvement and preventing the development of communication skills.

\par The concept of \textit{Cognitive Offloading} describes the process of reducing cognitive load by utilizing external tools or systems~\cite{risko2016cognitive,scaife1996external,armitage2020developmental}. \textit{Understood} aligns with this concept by providing real-time support mechanisms—such as summarization, suggestions, and off-topic detection—that help alleviate cognitive strain for users, particularly those with ADHD. These features allow users to focus on the conversation itself while offloading some of the cognitive demands. While this offloading helps users stay engaged and confident in the moment, excessive reliance on such features could hinder cognitive growth and self-regulation~\cite{grinschgl2021consequences}. Thus, it is crucial to strike a balance between providing cognitive support and fostering cognitive development. If support becomes too frequent or easily accessible, users may skip the cognitive effort required to engage deeply in conversations. Therefore, the system should provide temporary support that gradually fades as users become more adept at independent communication.

\par To address this balance, we implemented two key design strategies: (1) \textbf{Active Trigger Design}, where the support trigger is positioned outside the immediate field of view, encouraging users to actively seek assistance only when needed, and (2) \textbf{Time-Limited Support}, where suggestions disappear after a certain period of time, maintaining a clean interface and further preventing over-reliance. These design choices ensure that support remains optional and that users retain control over their engagement. By minimizing excessive reliance on the system and gradually fading support, \textit{Understood} aims to help users develop independent communication skills in the long term.

\par In conclusion, addressing the tension between \textit{Support} and \textit{Self-Reliance} requires providing support that reduces cognitive load while encouraging independent thinking. Future designs should prioritize dynamic balance, offering assistance that complements users' ability to independently navigate challenges, fostering both immediate help and long-term self-reliance.

\subsection{Safeguarding User's Autonomy through System Usage}
\label{sec:Discussion-Autonomy}

\par During our user study, concerns were raised about participants’ potential over-reliance on \textit{Understood}, which could lead to a diminished sense of autonomy. Autonomy refers to the ability to make self-directed decisions free from external coercion or normative pressure~\cite{wertenbroch2020autonomy, neurodivergentAutonomy, fink2024let}. Recent work has emphasized autonomy as a key factor influencing user trust and engagement~\cite{vantrepotte2022leveraging, pagliari2022new}. For example, Villa et al.~\cite{villa2024envisioning} found that autonomy can be enhanced by explicitly providing users with control and increasing their confidence during conversations.

\par \textit{Understood} offers real-time communication support such as summarization and suggestions, intended to help users express themselves more comfortably. However, we recognize that if users follow these suggestions passively, it may create a subtle power imbalance, leading them to defer to the system rather than reflect on their own communicative intent.

\par To address this concern, we emphasize that \textit{Understood} is not designed to prescribe any normative communication style. Instead, its functionality is grounded in customizability and user control. In the Function Selection Phase, users choose the types of support they wish to activate based on their personal needs. In the Conversation Phase, support is provided only through explicit user triggering, while visual cues fade over time to reduce intrusiveness and the risk of passive reliance on the system. Finally, the Feedback Phase allows users to reflect on their experience and opt out at any time, thereby avoiding pressure or normative imposition.

\par That said, while we acknowledge the autonomy risks inherent in any assistive technology, we believe that \textit{Understood} safeguards autonomy through intentional, reflexive, and user-centered design.

\subsection{Complementing Long-Term Root-Cause Interventions with Immediate Symptomatic Support}
\label{sec:Discussion-Balance}

\par In our user study, clinicians acknowledged the effectiveness of \textit{Understood} in helping individuals with ADHD navigate moment-to-moment communication challenges. However, they also expressed concerns about the need to balance short-term symptomatic relief with long-term, root-cause therapeutic strategies. \textit{Understood} offers immediate support mechanisms -- such as conversation summarization, word suggestions, and off-topic detection -- that primarily address the surface-level manifestations of ADHD. These features function like ``training wheels'' when learning to ride a bicycle: they compensate for current functional gaps but do not directly address the underlying cognitive causes of attention dysregulation. While it does not address the underlying causes, such symptomatic support can significantly reduce communication anxiety and social stigma, making it especially valuable for individuals who lack consistent access to structured, long-term treatment.

\par In contrast, interventions such as CBT and SST aim to achieve root-level change by targeting deeper cognitive mechanisms such as attention regulation, self-monitoring and strategy generation~\cite{willis2019stand,ladd1983cognitive,gokel2017effects}. These therapies rely on guided feedback, structured practice, and cognitive restructuring over extended periods. Using the same metaphor, while our system helps \textit{``hold the user balanced before they fall''}, traditional therapies aim to \textit{``teach the user how to maintain balance on their own''}.

\par To strike a balance between these two modes of support, all four auxiliary functions in \textit{Understood} are designed to be user-triggered, activated only when the user deliberately gazes downward at the trigger box. This design ensures that assistance is delivered only when actively sought, helping preserve cognitive engagement and user agency. Furthermore, we explicitly position \textit{Understood} as a complement -- not a replacement -- for therapist-mediated interventions. This distinction is crucial given the phenomenon of \textit{Context Dependence}~\cite{edwards2024real,el2019context}, where users may perform well within structured therapy environments but struggle to apply learned strategies in unstructured, real-world scenarios. In this light, \textit{Understood} may serve as a valuable daily-life support layer, bridging formal interventions and everyday application. Future work may explore how to further personalize in-situ support by aligning real-time system behavior with individualized therapeutic goals.

\subsection{Generalizing \textit{Understood} beyond ADHD}
\par Our user study demonstrates that \textit{Understood} effectively supports individuals with ADHD in overcoming communication difficulties, while preserving ease of use. Beyond ADHD, other neurodivergent populations -- such as individuals with aphasia, autism spectrum disorder (ASD), or pragmatic language impairment (PLI) -- often face analogous communicative challenges, including memory disruptions and conversational disfluency~\cite{CLARK2003265,MARKOWITSCH2003287,Wilkinson1998Profiles,Jenny2013Social}. These parallels suggest the potential for \textit{Understood} to be adapted to a broader range of user needs.

\par However, while shared difficulties exist, the nuanced nature of these conditions necessitates a tailored approach. Future adaptations should be grounded in an in-depth understanding of the specific cognitive and communicative profiles of each group. This would enable the derivation of refined design requirements to better accommodate diverse deficits and usage contexts.

\par Moreover, the applicability of \textit{Understood} may extend beyond the domain of neurological disorders. For example, second-language (L2) speakers often experience similar breakdowns in fluency and conversational flow due to cognitive overload or limited linguistic resources. In such cases, \textit{Understood} could act as an immersive communication scaffold -- providing predictive suggestions that support word retrieval, reduce cognitive load, and foster linguistic intuition. This aligns with the principle articulated by Elise Roy in her TED talk: \textit{``When we design for disability, we all benefit''}~\cite{EliseRoy}.

\subsection{Limitations and Future Work}
\par Our research has several limitations that warrant consideration. 

\paragraph{System Performance}
\par Our research is limited by several technological constraints that impact the overall performance of the system. First, the design space of the supportive system may not be fully explored. The initial design concepts were derived from an exploratory brainstorming process, which, while generating a broad range of ideas, may still overlook certain potential solutions. Future work could expand the design space by integrating a wider variety of approaches, enabling a more comprehensive exploration of support strategies. 

\par Additionally, although the \textit{HoloLens} provides a comprehensive heads-up display for delivering assistive functionalities, we acknowledge that such devices tend to be highly visible, potentially disrupting the social dynamics of face-to-face interactions. This visibility may increase the risk of stigmatization for our users with ADHD, reinforcing stereotypes such as being perceived as \textit{``difficult to communicate with''}. Future designs might consider alternative support platforms with lower visual intrusiveness, such as Google Glass, and could incorporate complementary modalities like subtle auditory cues to deliver assistance in a less obtrusive manner.

\par Third, the performance of \textit{Understood} is constrained by external technological limitations, specifically the use of the \textit{GPT-4o} model. We prioritized accurate feedback over response time, which led to occasional delays in generating responses. While alternatives like \textit{GPT-3.5 Turbo}~\cite{gpt3.5} and \textit{Phi-4-Mini}~\cite{phi4} offer quicker response times (around 1.2 seconds and 0.6 seconds, respectively), these models may compromise on the accuracy of the feedback, producing vague or overly lengthy responses. Despite the response delays, \textit{GPT-4o} provides more accurate feedback, which aligns with our system’s goals. Additionally, LLMs can still introduce hallucinations in feedback, as observed in our user study, where users noted issues with the perceived correctness of the responses. Future advancements in models like \textit{GPT o3-mini}~\cite{gpto3} and \textit{DeepSeek-R1}~\cite{deepseek} may offer more accurate responses, but at the cost of longer response times, which would not meet our real-time requirements. 

\par Finally, we relied on \textit{Azure} speech services to differentiate conversation participants, using a single-source microphone (HoloLens). This approach may be less accurate than using individual microphones, but such setups are costly and impractical. Future research should focus on addressing these technical issues to benefit underrepresented groups like individuals with ADHD. 

\paragraph{User Study}
\par Our user study has several limitations that should be addressed in future research. First, the sample size of \textit{N=10}, consisting primarily of university students with ADHD, is relatively small. ADHD remains a minority condition, and recruiting participants from this group is inherently challenging, which limited the sample size in our study. Consequently, the sample may not fully represent the broader adult ADHD population. Future work could include a more diverse range of participants across different age groups to capture a wider variety of needs and experiences. 

\par Additionally, our study adapted items from the NASA-TLX and related tools to assess perceived mental/physical demand, performance, and frustration, complemented by objective indicators such as response length and accuracy. These jointly support the claim that the system facilitates cognitive offloading. However, as the full NASA-TLX was not directly administered, this may limit the completeness and comparability of our assessment. We acknowledge this limitation and suggest that future work incorporate the full instrument for more standardized evaluation.

\par It's also worth noting that our study did not compare \textit{Understood} with traditional interventions, such as SST and CBT, which are long-term approaches. Instead, \textit{Understood} was positioned as a complementary tool for individuals who face challenges accessing these traditional therapies. Future work could explore how \textit{Understood} can function as a supplement to, or even as a part of, longer-term interventions. 

\par Fourth, we also acknowledge the lack of long-term evaluations. Limited by time constraints, we did not assess the users’ experiences or progress over an extended period. Future research could incorporate longitudinal studies to better understand how users’ performance and experiences evolve over time. 

\par Finally, our study focused on a researcher-participant-only scenario to minimize bias, but it did not comprehensively evaluate the perceptions of conversation partners. Future research could design studies that include a broader range of perspectives, including those of conversation partners, to gain a more holistic understanding of the system's impact. 

\paragraph{Future Design Opportunities}
\par As \textit{Understood} evolves, several design opportunities can enhance its effectiveness. A key challenge lies in balancing \textit{Support} and \textit{Self-Reliance}. While real-time support reduces cognitive load, it’s essential to avoid over-reliance that hinders independent thinking. Future designs should refine current strategies such as Active Trigger Design and Time-Limited Support, ensuring support fades gradually, encouraging users to become more self-reliant over time. 

\par Another opportunity is in facilitating the transfer of skills from \textit{Understood} to users' daily routines. Adaptive features could guide users in applying communication strategies in real-world contexts, ensuring long-term cognitive growth. Furthermore, the system should continue to complement traditional therapies, like CBT and SST, by aligning real-time support with individualized therapeutic goals, bridging structured interventions and everyday application. 

\par Meanwhile, MR platforms, such as \textit{HoloLens}, offer significant design potential, but their high cost limits accessibility. Future designs could explore transitioning \textit{Understood} to more affordable MR devices or even smartphones, broadening the system’s reach while maintaining immersive support, although some features may be discarded. 

\par Finally, \textit{Understood} could be adapted to support other neurodivergent populations and even second-language learners, addressing common communication challenges. Tailoring the system to these groups could extend its reach and enhance its utility.

\section{Conclusion}
\par In this study, we introduced \textit{Understood}, an MR system designed to assist adults with ADHD in real-world communication. Developed through a formative study with semi-structured interviews and a design study, \textit{Understood} features three key features on (1) real-time conversation summarization, (2) context-aware subsequent word suggestions and (3) topic shifting detection and reminding. User study demonstrated that , when evaluated under a within-subject, lab-based setting, \textit{Understood} effectively supports communication with high usability, offering a complement to therapist-mediated interventions. This research not only deepens our understanding of the communication challenges faced by adults with ADHD, but also prompts a dialectical reflection on the balance between external support and self-reliance, as well as the interplay between symptomatic support and root-cause interventions. We view \textit{Understood} as a step towards advancing communication practices, potentially paving the way for more sophisticated human-computer collaboration in accessibility contexts.

\begin{acks}
\par We thank anonymous reviewers for their valuable feedback. We would like to thank Dr. Sichu Wu at The Affiliated Brain Hospital of Nanjing Medical University, Dr. Hao Yao at Shanghai Mental Health Center and Yuchen Wu from University of Waterloo for their insightful suggestions. We would like to express our gratitude to \textit{Irasutoya}\footnote{\url{https://www.irasutoya.com/}} for inspiring the illustration style used in part of this paper. This work is supported by grants from the National Natural Science Foundation of China (No. 62372298), Shanghai Engineering Research Center of Intelligent Vision and Imaging, Shanghai Frontiers Science Center of Human-centered Artificial Intelligence (ShangHAI), and MoE Key Laboratory of Intelligent Perception and Human-Machine Collaboration (KLIP-HuMaCo).
\end{acks}

\bibliographystyle{ACM-Reference-Format}
\bibliography{sample-base}

\onecolumn
%TC:ignore
\appendix

\section{15 Statements of Communication Challenges}
\label{app:DIVA-5}

\begin{table}[h]
\begin{tabular}{c|c|c}
\textbf{Statements of Communication Challenges}                   & \textbf{\begin{tabular}[c]{@{}c@{}}Ratings by Clinicians\\ (Sorted)\end{tabular}} & \textbf{\begin{tabular}[c]{@{}c@{}}Occurrences on Individuals with ADHD\\ (Counts)\end{tabular}} \\ \hline
1. I do not like reading due to mental effort.                    & 2, 2, 2, 3                                                                        & 6                                                                                   \\
2. I have difficulty on speaking fluently in a long sentence.     & 2, 3, 3, 3                                                                        & 10                                                                                  \\
3. I have difficulty concentrating on a conversation.             & 2, 3, 3, 3                                                                        & 9                                                                                   \\
4. I often change the subject of the conversation.                & 2, 3, 3, 3                                                                        & 8                                                                                   \\
5. I easily get distracted by the conversations of others.        & 2, 2, 3, 3                                                                        & 7                                                                                   \\
6. I have difficulty in filtering and/or selecting information.   & 3, 3, 3, 3                                                                        & 9                                                                                   \\
7. I talk during activities when this is not appropriate.         & 2, 2, 3, 3                                                                        & 6                                                                                   \\
8. I have difficulty in speaking softly.                          & 1, 2, 2, 3                                                                        & 5                                                                                   \\
9. I have a tendency to talk too much.                            & 2, 2, 3, 3                                                                        & 8                                                                                   \\
10. I do not give others room to interject during a conversation. & 2, 2, 3, 3                                                                        & 6                                                                                   \\
11. I need a lot of words to say something.                       & 2, 2, 3, 3                                                                        & 8                                                                                   \\
12. I have difficulty waiting my turn during conversations.       & 3, 3, 3, 3                                                                        & 11                                                                                  \\
13. I interrupt others frequently.                                & 2, 3, 3, 3                                                                        & 8                                                                                   \\
14. I give people answers before they have finished speaking.     & 2, 3, 3, 3                                                                        & 9                                                                                   \\
15. I say things without thinking first.                          & 2, 3, 3, 3                                                                        & 9                                                                                  
\end{tabular}
\caption{The 15 statements derived from DIVA-5 regarding the communication challenges on individuals with ADHD, with the ratings by clinicians and occurrences on individuals with ADHD for each statement.}
\label{tab:DIVA-5}
\end{table}

\section{Summarized Storyboards in Design Workshop}
\label{app:allStoryboards}
\begin{figure}[h]
    \centering
    \includegraphics[width=0.65\linewidth]{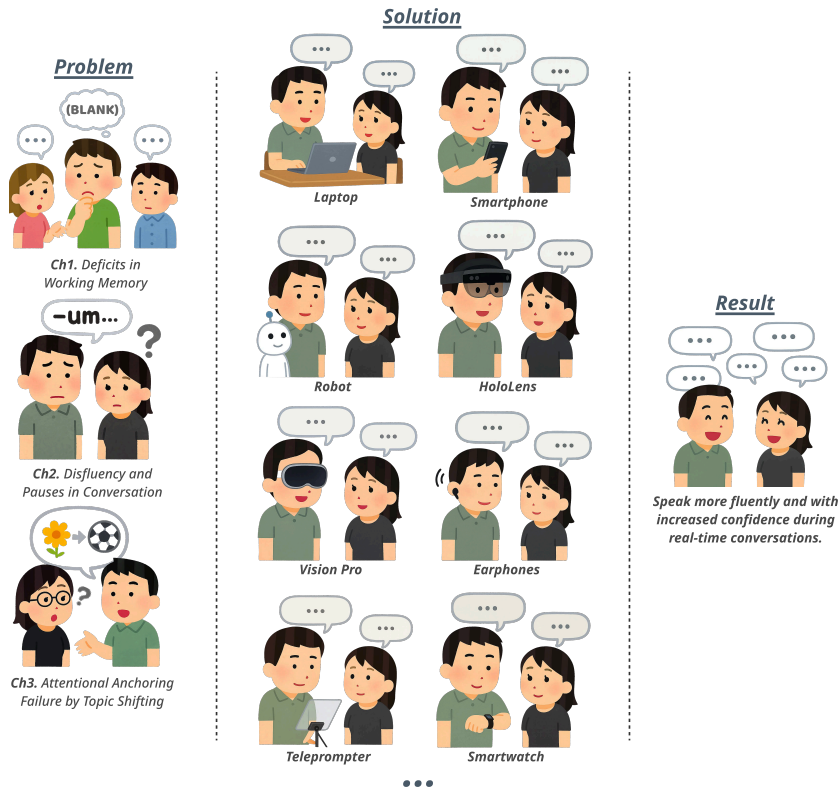}
    \caption{Summarized version of the storyboards used in our design workshop.}
    \label{fig:allStoryboards}
\end{figure}

\section{Prompt Templates}
\subsection{Summarization}
\label{app:SummarizationPrompt}
\par \texttt{You are a large language model assistant specialized in dialogue understanding and summarization. Your task is to generate a 4 to 12 words concise, high-level summary of the given recent utterance and previous summary by extracting the most relevant keywords that capture the core meaning or topic of the dialogue.}

\par \texttt{Instructions:}
\par \texttt{- Read the recent utterance and the previous summary provided below.}
\par \texttt{- The keywords should be semantically meaningful and representative of the dialogue content.}
\par \texttt{- Stay within the character limit strictly.}

\par \texttt{Input Fields:}
\par \texttt{- Recent Utterance}: \textit{<recent utterance>}
\par \texttt{- Previous Summary}: \textit{<previous summary>}

\subsection{Word Suggestions}
\label{app:WordSuggestionsPrompt}
\par \texttt{You are a large language model assistant specialized in predictive text generation. Your task is to help a human by predicting the next plausible phrase based on the current utterance and the full dialogue history. You are not generating a full sentence but only a short phrase or a set of words less than 6 words. It must not exceed this length, and it must not be a complete sentence. The predicted phrase should naturally follow the given context, as if continuing the flow of the conversation.}

\par \texttt{Instructions:}
\par \texttt{- Read the current utterance and the full dialogue history provided below.}
\par \texttt{- Do not include a full sentence.}
\par \texttt{- Stay within the character limit strictly.}

\par \texttt{Input Fields:}
\par \texttt{- Current Utterance}: \textit{<current utterance>}
\par \texttt{- Dialogue History}: \textit{<full dialogue history>}

\subsection{Off-topic Detection}
\label{app:Off-topicDetectionPrompt}
\par \texttt{You are a large language model assistant tasked with evaluating topic consistency within a conversation. Given the current utterance and the full dialogue history, your task is to determine whether the current utterance remains on topic or deviates from the main thread of the conversation. You should report ``Yes'' if the utterance deviates from the topic, or ``No'' if it remains on topic.}

\par \texttt{Instructions:}
\par \texttt{- Read the current utterance and the full dialogue history provided below.}
\par \texttt{- The judgment should be based solely on the current utterance and prior conversation.}
\par \texttt{- Do not generate any content other than ``Yes'' or ``No''.}

\par \texttt{Input Fields:}
\par \texttt{- Current Utterance}: \textit{<current utterance>}
\par \texttt{- Dialogue History}: \textit{<full dialogue history>}

\newpage
\section{Details of the User Study}
\subsection{Open-ended Conversation Topics and Supplementary Information}
\label{app:userstudy-topic}
\begin{enumerate}
    \item \textbf{Topic:} What is your favorite place in the city you currently live in?
        \par \textbf{Supplement:} \textit{As for me, my favorite place in the city is Riverside Park. I really enjoy going there in the late afternoon, especially on weekdays when it's quieter. There’s a little bench under a large maple tree where I like to sit and read or just relax. The view of the river is really calming, and sometimes you can see ducks swimming by. I also like that it's not too far from my apartment -- just a ten-minute walk. It’s the kind of place that helps me reset after a busy day.}
    \item \textbf{Topic:} Can you describe the things you usually do on weekends?
        \par \textbf{Supplement:} \textit{One thing I often do on weekends is visit a small coffee shop near my apartment called Blue Bean. I usually go there around 2 p.m. on Saturdays to meet a friend. We both like trying different kinds of coffee, and they also serve really good almond croissants. Last weekend I tried their seasonal lavender latte -- it was surprisingly good! The place has a quiet upstairs area where we usually sit and talk for an hour or two. It’s become kind of a routine for us, and I always look forward to it.}
    \item \textbf{Topic:} What is your go-to meal when you're busy or tired?
        \par \textbf{Supplement:} \textit{When I’m really tired or just don’t have much time, my go-to meal is definitely egg fried rice. It’s quick to make and I usually have all the ingredients at home -- just leftover rice, an egg, some chopped green onions, and soy sauce. Sometimes I add frozen peas or bits of ham if I have them. I make it in one pan, which also means less cleanup. I’ve been making it since college, so I don’t even have to think about the steps anymore -- it’s almost automatic. It’s simple but always satisfying, especially on a busy weekday evening.}
    \item \textbf{Topic:} What’s your favorite way to relax after a long day?
        \par \textbf{Supplement:} \textit{After a long day, my favorite way to relax is making a cup of herbal tea and watching a light TV show. I usually drink chamomile or peppermint tea -- something without caffeine. I turn off most of the lights and just leave a small lamp on in the living room to make it feel cozy. Then I put on something easy to watch, like a cooking show or a feel-good sitcom. I try not to check my phone during that time so I can fully unwind. It’s a little routine that helps me mentally switch off and feel calmer before bed.}
    \item \textbf{Topic:} What apps do you use most on your phone, and what for?
        \par \textbf{Supplement:} \textit{The app I use the most is probably Notion. I use it to keep track of my to-do lists, project notes, and even meal plans. I usually check it first thing in the morning to review my schedule, and then again in the evening to update what I’ve done. I have different pages set up -- for example, one for work tasks, one for personal goals, and another for weekly reflections. It syncs across my phone and laptop, which makes it super convenient. Without it, I’d probably forget half the things I need to do.}
\end{enumerate}

\subsection{Questionnaire Design}
\label{app:userstudy-question}
\subsubsection{Effectiveness}
\par We measured ten aspects using 5-point Likert scale (1 = strongly disagree, 5 = strongly agree):
\begin{enumerate}
    \item \textbf{Response Length (Summarization):} ``I felt that the length of the summarization was appropriate.''
    \item \textbf{Response Length (Suggestion):} ``I felt that the length of the suggestion for subsequent words was appropriate.''
    \item \textbf{Response Time (Summarization):} ``I felt that the time taken for summarization was acceptable.''
    \item \textbf{Response Time (Suggestion):} ``I felt that the time taken for suggestion for subsequent words was acceptable.''
    \item \textbf{Correctness (Summarization):} ``I felt that the system correctly summarized the conversation.''
    \item \textbf{Correctness (Suggestion):} ``I felt that the system correctly suggested the subsequent words.''
    \item \textbf{Concentration on Conversation:} ``I was always concentrating on the conversation.''
    \item \textbf{Eye Contact:} ``When I was speaking, I maintained eye contact with partners.''
    \item \textbf{Naturalness:} ``I acted naturally at all times during the conversation.''
    \item \textbf{Helpfulness:} ``I felt that the system helped me overcome the challenges in the conversation.''
\end{enumerate}

\subsubsection{Usability}
\par We measured ten aspects using 5-point Likert scale (1 = strongly disagree, 5 = strongly agree):
\begin{enumerate}
    \item \textbf{Ease of Use:} ``I felt that the system was easy to handle.''
    \item \textbf{Learnability:} ``I felt that I can quickly learn how to use the system.''
    \item \textbf{Function Integration:} ``I felt the functions in the system are well-integrated.''
    \item \textbf{User Control:} ``I felt that the system was under my control and behaved as I expected.''
    \item \textbf{Low Interruption:} ``The system would seldom interrupt the conversation.''
    \item \textbf{Low Distraction:} ``I felt that there was little distraction in the system design.''
    \item \textbf{Low Mental Demand:} ``I felt that it was not mentally demanding to use the system.''
    \item \textbf{Low Physical Demand:} ``I felt that it was not physically demanding to use the system.''
    \item \textbf{Satisfaction:} ``I felt satisfied after using the system to conduct conversation.''
    \item \textbf{Frequent Usage:} ``I would like to use the system frequently.''
\end{enumerate}

%TC:endignore
\end{document}